\documentclass[a4paper]{article}
\usepackage[margin=25mm]{geometry}
\usepackage{url}
\usepackage{import}
\usepackage{amsfonts,amssymb,amsmath,amsthm,dsfont}
\usepackage{xcolor}
\usepackage{comment}
\usepackage[linesnumbered,algoruled]{algorithm2e}
\usepackage{bm}
\usepackage[normalem]{ulem}
\usepackage{graphicx}
\usepackage{adjustbox}
\usepackage{subcaption}

\definecolor{light-gray}{gray}{0.75}

\newcommand{\ve}[1]{ {\mathbf{#1}} }
\usepackage{mathtools}
\usepackage{eurosym}
\DeclarePairedDelimiter\floor{\lfloor}{\rfloor}

\begin{document}

\title{Phase retrieval with Bregman divergences and application to audio signal recovery\thanks{This work is supported by the European Research Council (ERC FACTORY-CoG-6681839).}}
\date{}
\author{Pierre-Hugo~Vial\thanks{IRIT, Université de Toulouse, CNRS, Toulouse, France (e-mail: firstname.lastname@irit.fr).} ,
        Paul~Magron\footnotemark[2] ,
        Thomas~Oberlin\thanks{ISAE-SUPAERO, Université de Toulouse, France (e-mail: firstname.lastname@isae-supaero.fr).} ,
        Cédric~Févotte\footnotemark[2]
}


\maketitle

\begin{abstract}
Phase retrieval (PR) aims to recover a signal from the magnitudes of a set of inner products. This problem arises in many audio signal processing applications which operate on a short-time Fourier transform magnitude or power spectrogram, and discard the phase information. Recovering the missing phase from the resulting modified spectrogram is indeed necessary in order to synthesize time-domain signals. PR is commonly addressed by considering a minimization problem involving a quadratic loss function. In this paper, we adopt a different standpoint. Indeed, the quadratic loss does not properly account for some perceptual properties of audio, and alternative discrepancy measures such as beta-divergences have been preferred in many settings. Therefore, we formulate PR as a new minimization problem involving Bregman divergences. Since these divergences are not symmetric with respect to their two input arguments in general, they lead to two different formulations of the problem. To optimize the resulting objective, we derive two algorithms based on accelerated gradient descent and alternating direction method of multipliers. Experiments conducted on audio signal recovery from spectrograms that are either exact or estimated from noisy observations highlight the potential of our proposed methods for audio restoration. In particular, leveraging some of these Bregman divergences induce better performance than the quadratic loss when performing PR from spectrograms under very noisy conditions.
\end{abstract}

\section{Introduction}
\label{sec:intro}

Data reconstruction from phaseless measurements is a problem that arises in various fields including X-ray crystallography \cite{harrison93}, optics \cite{walther63} and astronomy \cite{fienup87}. This task, hereafter termed phase retrieval (PR), is also ubiquitous in audio signal processing, where much research has focused on the processing of nonnegative time-frequency representations such as short-time Fourier transform (STFT) magnitude or power spectrograms. Processing STFT spectrograms results in discarding or not accounting for the phase information: it is then necessary to retrieve the missing phase in order to synthesize time-domain signals. Therefore, PR is of paramount importance for tasks that involve audio signal reconstruction from incomplete time-frequency observations. Consequently, it has attracted some attention for many applications such as speech enhancement~\cite{Krawczyk2012,Gerkmann2015,Mowlaee2016}, source separation~\cite{Magron2018,Wang2018a,Wichern2018,Wang2019a} or audio restoration (e.g., click removal~\cite{Magron2015a} or time-frequency inpainting~\cite{Kreme2018}).  \par
PR consists in recovering a signal $\mathbf x^\star \in \mathbb C ^L$ from nonnegative measurements $\mathbf r \approx |\mathbf A \mathbf x^\star|^d \in \mathbb R_+^K$, where $\mathbf A \in \mathbb C^{K\times L}$ is the measurement matrix and $d$ is usually equal to $1$ or $2$, depending whether one considers magnitude or power measurements. This problem is inherently ill-posed as different signals can generate identical measurements. Thus, $\mathbf x^\star$ can only be recovered up to several ambiguities which depend on $\mathbf A$. In particular, the STFT magnitude of a considered signal cannot uniquely represent this signal without specific constraints or a priori knowledge about part of the samples \cite{nawab1983signal}. For example, estimation is subject to a global phase ambiguity, as the magnitude spectrograms of $\mathbf x^\star$ and $c\mathbf x^\star$ are identical when $c \in \mathbb C$ and $|c|=1$. PR is commonly formulated as a nonconvex minimization problem involving a quadratic loss function, as follows:
\begin{equation}
    \label{eq:pr}
    \underset{\mathbf x \in \mathbb C^L}{\text{min}}\quad E(\mathbf x):= \|\mathbf r - |\mathbf A \mathbf x|^d\|_2^2 .
\end{equation}
Problem~\eqref{eq:pr} may be tackled with conventional optimization algorithms such as gradient descent \cite{WF}, \cite{decorsiere2014inversion}, alternating projections \cite{GSA, fienup82}, majorization-minimization \cite{PRIME}, alternating direction method of multipliers (ADMM) \cite{wen12, liang17}, and leveraging the structure of time-frequency measurements~\cite{Bendory2018,Pfander2019}. An extensive review of those algorithms from a numerical perspective can be found in \cite{fannjiang2020numerics}. Convex optimization approaches are also considered in \cite{candes2013phaselift, candes2013phase, waldspurger2015phase, jaganathan2015recovering} by lifting the problem to a higher dimensional space (i.e., solving a constrained quadratic problem involving $ \mathbf x \mathbf x^\mathsf H$) and relaxing the rank-one constraint. However, they are impracticable for processing audio signals, as they square the dimensionality of the problem~\cite{Sun2012}. The Griffin-Lim algorithm (GLA) \cite{GLA}, a variant of the Gerchberg-Saxton algorithm (GSA) \cite{GSA} adapted to STFT measurements, is one of the most popular techniques in the audio literature and is generally considered as a baseline for signal recovery. PR has also been tackled using signal modeling~\cite{Magron2015a,Prusa2017,Prusa2017spl} or deep neural networks~\cite{Takamichi2018}. However, optimization-based approaches remain efficient, provide theoretical guarantees and may still be used with model-based approaches~\cite{Masuyama2019deepGLA,Masuyama2021}.

\par Even though a considerable amount of research has been conducted to tackle the PR problem as described in~\eqref{eq:pr}, such an approach suffers from one drawback when it comes to audio. Indeed, it is well established that the quadratic loss is not the best-suited metric for evaluating discrepancies in the time-frequency domain. For instance, it does not properly characterize the perceptually-related properties of audio such as its large dynamic range~\cite{gray80}. \par
As such, in this work we propose to replace the quadratic loss function in~\eqref{eq:pr} by alternative divergences which are more appropriate for audio signal processing~\cite{gray80}. We consider general Bregman divergences, a family of loss functions which encompasses the beta-divergence~\cite{cic10,Hennequin2011} and some of its well-known special cases, the generalized Kullback-Leibler (KL)\footnote{We will simply term it ``KL divergence" in this paper for brevity.} and Itakura-Saito (IS) divergences. These are acknowledged for their superior performance in nonnegative audio spectral decomposition \cite{Virtanen2007, fevotte09, smaragdis14}, \cite{Magron2018iwaenc}, audio inpainting~\cite{LeRoux2011}, and music analysis~\cite{Hennequin2011a, Vincent2010}. Besides, these divergences naturally arise from a statistical perspective. For instance, minimizing the KL divergence between an observed spectrogram and a parametric one assumes that the observations follow a Poisson model. Similarly, minimizing the IS divergence implies a multiplicative Gamma noise model~\cite{smaragdis14}. In order to be as general as possible, we consider any nonnegative power $d$ (we do not restrict to either $1$ nor $2$) and we account for the fact that these divergences are not symmetric with respect to their input parameters in general, which actually leads to tackling two different problems. To optimize the resulting objective, we derive two algorithms, based on accelerated gradient descent~\cite{NESTEROV1987} and ADMM \cite{boyd11}. We experimentally assess the potential of our approach for PR on music and speech restoration tasks. Our experimental results show that some of the proposed methods compare favorably or outperform traditional methods based on the quadratic loss. In particular, the proposed gradient algorithm based on the KL divergence or the beta-divergence with $\beta=0.5$ seems promising when the spectrogram is corrupted, showing more robustness than the algorithms based on quadratic loss.

The rest of the paper is organized as follows. Section~\ref{sec:related_work} reviews several baseline algorithms for PR. Section~\ref{sec:pr_bregman} describes the PR problem extended to Bregman divergences and the two proposed algorithms. Section~\ref{sec:exp} presents the experimental results for audio signal recovery applications. Finally, Section~\ref{sec:conclusion} draws some concluding remarks. For the sake of generality, we assume $\mathbf{x}$ to be complex-valued everywhere in Sections~\ref{sec:related_work} and \ref{sec:pr_bregman}. Transposition to the real-valued case is discussed in Section~\ref{sec:exp} and in Appendix~\ref{sec:apdx_real}.

\vspace{0.5em}
\noindent \textbf{Mathematical notations}:
\begin{itemize}
    \item $\mathbf{A}$ (capital, bold font): matrix, whose $(m,n)$-th entry is denoted $A(m,n)$.
    \item $\mathbf{x}$ (lower case, bold font): column vector, whose $t$-th entry is denoted $x(t)$.
    \item $z$ (regular): scalar.
    \item $|.|$, $\angle(.)$, $(.)^*$: magnitude, complex angle, and complex conjugate, respectively.
    \item $^\mathsf{T}$, $^\mathsf{H}$: transpose and Hermitian transpose, respectively.
    \item $\mathfrak R$, $\mathfrak I$: real and imaginary part functions.
    \item $\odot$, $(.)^d$, fraction bar: element-wise matrix or vector multiplication, exponentiation, and division, respectively.
    \item $\mathbf{I}_L$: identity matrix of size $L$.
    \item $\mathcal P_\mathcal S$ : projection operator on the set $\mathcal S$.
\end{itemize}

\section{Related work}

\label{sec:related_work}

In this section, we present three state-of-the-art approaches related to our own contributions: alternating projections (Section~\ref{sec:2A}), gradient descent (Section~\ref{sec:2B}), and ADMM (Section~\ref{sec:2C}). Note that PR being a non-convex optimization problem, the outputs of the descent methods considered in this paper are influenced by the initialization.

\subsection{Alternating projections}
\label{sec:2A}

In the seminal work~\cite{GLA}, the authors address the PR problem \eqref{eq:pr} with $d=1$ and with $\mathbf A$ being the STFT operator. They propose to alternate projections on $\mathcal M$, the set of time-frequency coefficients whose magnitude is equal to the observed measurements, and $\mathcal C$, the set of \emph{consistent} coefficients, that is, complex coefficients that correspond to the STFT of time-domain signals~\cite{leroux08}. More formally, we have:
\begin{equation}
    \mathcal M = \{ \tilde{\mathbf x} \in \mathbb C^K \mbox{ } |  \mbox{ } |\tilde{\mathbf x}| = \mathbf r \}
    \mbox{ and } \mathcal C = \{ \tilde{\mathbf x} \in \mathbb C^K \mbox{ } |  \mbox{ } \tilde{\mathbf x} = \mathbf A \mathbf A^\dag\tilde{\mathbf x} \},
    \label{eq:proj_spaces}
\end{equation}
where $\tilde{\mathbf x}$ is a vector of complex-valued time-frequency coefficients and ${\mathbf A^\dag = (\mathbf A^\mathsf H \mathbf A)^{-1} \mathbf A^\mathsf H}$ is the Moore-Penrose pseudo-inverse of $\mathbf A$ (which encodes the inverse STFT), and where here $^{-1}$ denotes the matrix inverse. When the window used in the STFT is self-dual (i.e., it leads to perfect reconstruction if used for both analysis and synthesis), we have $\mathbf A^\mathsf H \mathbf A = \ve{I}_L$ and as such $\mathbf A^\dag = \mathbf A^\mathsf{H}$ (see Appendix \ref{sec:apdx_stft} for more details about the STFT and duality). We make such an assumption throughout the paper (without loss of generality). The two projections then write:
\begin{equation}
\label{eq:glaproj}
    \mathcal P_\mathcal M(\tilde{\mathbf x})=\mathbf r \odot \frac{\tilde{\mathbf x}}{|\tilde{\mathbf x}|} \mbox{ and }
    \mathcal P_\mathcal C(\tilde{\mathbf x})=\mathbf A \mathbf A^\mathsf H \tilde{\mathbf x}.
\end{equation}
Although $\mathcal  M$ is not a subspace and is not convex, we still call $\mathcal P_\mathcal M$ a projection since it maps an element of $\mathbb C^K$ to its closest element in $\mathcal M$ (in the mean squared error sense), which is unique~\cite{bauschke2002phase,FGLA} when $\mathcal{M}$ is defined as in~\eqref{eq:proj_spaces}. Alternating these projections results in GLA, which is proved to converge to a critical point of the quadratic loss in \eqref{eq:pr}~\cite{GLA}. Alternatively, this algorithm can also be obtained by majorization-minimization~\cite{PRIME}.

In \cite{FGLA}, an accelerated version of GLA, termed Fast GLA (FGLA), is proposed with a Nesterov-like scheme with constant acceleration parameter. FGLA was shown experimentally to reach lower local minima of the problem \eqref{eq:pr} with $d=1$, yet without theoretical convergence guarantee. Other improvements of GLA include real-time purposed versions~\cite{RTISI,RTISILA} and its extension to multiple signals for source separation~\cite{Gunawan2010,Magron2020}.

GLA is similar to GSA \cite{GSA} as they are both alternating projection algorithms. They yet differ in that GSA uses the discrete Fourier transform (DFT) as the measurement operator and accounts for an additional constraint on the support of the time-domain signal to make the solution unique. For GLA, this constraint is not necessary as uniqueness can be obtained thanks to the redundancy of the STFT~\cite{Jaganathan2016}.

\subsection{Gradient descent}
\label{sec:2B}

In~\cite{WF}, Candès et al. address the PR problem~\eqref{eq:pr} with power measurements (i.e., $d=2$) and a general measurement matrix $\mathbf A$ (such as Gaussian random or DFT vectors). They propose to minimize the error $E$ with a gradient method. As the objective function implies complex quantities but is not holomorphic (i.e., not complex-differentiable), the authors express the gradient using the Wirtinger formalism~\cite{Bouboulis2010} detailed in Appendix~\ref{sec:apdx_wirtinger}. This leads to:
\begin{equation}
    \nabla E(\mathbf x) = \mathbf A^\mathsf H [(\mathbf A\mathbf x) \odot (|\mathbf A \mathbf x|^2 - \mathbf r)].
\end{equation}
The gradient algorithm update then writes:
\begin{equation}
    \mathbf x_{t+1}=\mathbf x_t - \mu_{t+1}\nabla E(\mathbf x_t),
\end{equation}
where $t$ is the iteration index and $\mu_t$ stands for the step size at iteration $t$. This approach is called the Wirtinger flow algorithm~\cite{WF}.

\subsection{ADMM}
\label{sec:2C}

\subsubsection{Minimization problem}

In \cite{liang17}, Liang et al. express PR as the following constrained problem by introducing auxiliary variables for the magnitude and phase of $\mathbf{A}\mathbf{x}$:
\begin{equation}
\label{eq:praux}
    \underset{\mathbf x \in \mathbb C^L, \mathbf u \in \mathbb R^K_+, \bm \theta \in \left [0;2 \pi \right [^K}{\text{min}}\, \|\mathbf r - \mathbf u\|^2_2 \mbox{  s.t. } \mathbf A \mathbf x = \mathbf u \odot e^{i \bm \theta},
\end{equation}
which is equivalent to~\eqref{eq:pr} when $d=1$. From~\eqref{eq:praux} one can derive the augmented Lagrangian:
\begin{equation}
     \mathcal L(\mathbf x, \mathbf u, \bm \theta, \bm \lambda)= \|\mathbf r - \mathbf u\|^2_2 + \mathfrak R \left( \bm \lambda^{\mathsf H} (\mathbf A\mathbf x-\mathbf u \odot e^{i\bm \theta}) \right)
     +\frac{\rho}{2}\|\mathbf A\mathbf x-\mathbf u \odot e^{i\bm \theta}\|_2^2 ,
\end{equation}
where $\bm \lambda$ is the vector of the Lagrange multipliers corresponding to the constraint $\mathbf A\mathbf x=\mathbf u \odot e^{i\bm \theta}$ and $\rho$ is the penalty parameter. From this expression, the authors derive the following ADMM update rules:
\begin{align}
    \{\mathbf u_{t+1}, \bm \theta_{t+1}\}&=\underset{\mathbf u\geq0, \bm \theta}{\text{argmin}}\, \mathcal L(\mathbf x_t, \mathbf u, \bm \theta, \bm \lambda_t), \label{eq:pr_admm_quad_1} \\
    \mathbf x_{t+1}&=\underset{\mathbf x}{\text{argmin}}\, \mathcal L(\mathbf x, \mathbf u_{t+1}, \bm \theta_{t+1}, \bm \lambda_t), \label{eq:pr_admm_quad_2} \\
    \bm \lambda_{t+1} &= \bm \lambda_t + \mathbf A\mathbf x_{t+1}-\mathbf u_{t+1}\odot e^{i \bm \theta_{t+1}}.
\end{align}
The closed-form expressions of~\eqref{eq:pr_admm_quad_1} and~\eqref{eq:pr_admm_quad_2} are provided in~\cite{liang17}. This procedure forms a special case of our proposed algorithm presented in Section~\ref{sec:3C}.

\subsubsection{Feasibility problem}
In \cite{wen12}, Wen et al. address PR with DFT measurements as a feasibility problem. Instead of~\eqref{eq:pr}, they consider the following formulation:
\begin{equation}
    \text{find}\quad \mathbf x \in \mathbb C^L \quad\text{s.t.}\quad \mathbf x \in \mathcal S_\mathcal F \cap  \mathcal S_0,
\end{equation}
where $ \mathcal S_\mathcal F$ is the set of signals whose DFT magnitude is $\mathbf{r}$ and $\mathcal S_0$ is the set of signals respecting an additional constraint (in optics, a typical constraint is that the signal is real-valued and nonnegative). They derive the following ADMM updates:
\begin{equation}
    \begin{aligned}
    \mathbf x_{t+1}&=\mathcal P_{\mathcal S_0}(\mathbf q_t - \mathbf p_t),\\
    \mathbf q_{t+1}&=\mathcal P_{\mathcal S_\mathcal F}(\mathbf x_{t+1}+\mathbf p_t), \\
    \mathbf p_{t+1} &= \mathbf p_t + \rho(\mathbf x_{t+1}-\mathbf q_{t+1}).
    \end{aligned}
    \label{eq:admm_fourier}
\end{equation}
The authors also note that for a penalty parameter $\rho=1$, this algorithm is equivalent to the hybrid input-output algorithm, which is well-known in optics~\cite{fienup82}. \par
In a similar fashion, Masuyama et al. \cite{Masuyama2019} use ADMM to tackle PR with STFT measurements (like in GLA) as a feasibility problem:
\begin{equation}
\text{find}\quad \tilde{\mathbf{x}} \in \mathbb C^K \quad\text{s.t.}\quad \tilde{\mathbf{x}} \in \mathcal M \cap  \mathcal C.
\end{equation}
Note that similarly as in Section~\ref{sec:2A}, we use the notation $\tilde{\mathbf{x}}$ to highlight that this approach operates in the (complex-valued) time-frequency domain.
They derive the following updates:
\begin{equation}
    \begin{aligned}
    \tilde{\mathbf x}_{t+1}&=\mathcal P_{\mathcal M}(\tilde{\mathbf q}_t - \tilde{\mathbf p}_t),\\
    \tilde{\mathbf q}_{t+1}&=\mathcal P_{\mathcal C}(\tilde{\mathbf x}_{t+1}+\tilde{\mathbf p}_t), \\
    \tilde{\mathbf p}_{t+1} &= \tilde{\mathbf p}_t + \tilde{\mathbf x}_{t+1}-\tilde{\mathbf q}_{t+1}.
    \end{aligned}
    \label{eq:admm_stft}
\end{equation}
This algorithm will be referred to as GLADMM. One can note than when $\mathbf p$ and $\tilde{\mathbf p}$ are equal to $0$, the algorithms defined by~\eqref{eq:admm_fourier} and~\eqref{eq:admm_stft} are respectively equivalent to GSA and GLA.

\section{Proposed methods}
\label{sec:pr_bregman}

In this section, we first propose a generalization of problem \eqref{eq:pr} to the family of Bregman divergences (Section \ref{sec:3A}). Then, relying on some of the related works presented in Section~\ref{sec:related_work}, we derive two algorithms based on accelerated gradient descent (Section \ref{sec:3B}) and ADMM (Section \ref{sec:3C}).

\subsection{Phase retrieval with general Bregman divergences}\label{sec:3A}

\begin{table*}[t]
    \normalsize
    \centering
    \caption{Typical Bregman divergences generating functions with their first and second derivatives. The KL and IS divergences are limit cases of the beta-divergence for $\beta = 1$ and $\beta=0$, respectively. The quadratic loss is obtained for $\beta=2$.}
    \label{tab:bregdiv}
    \begin{adjustbox}{scale=0.9}
    \begin{tabular}{c|cccc}
    \hline
    \hline
        Divergence & $d_{\psi}(y|z)$ & $\psi(z)$ &   $\psi'(z)$&  $\psi''(z)$ \\ \hline \hline
        Quadratic loss & $\frac{1}{2}(y-z)^2$  &  $\frac{1}{2}z^2$ &  $ z$ & $1$  \\
        Kullback-Leibler & $y(\log y - \log z) - (y-z) $ & $ z \log  z$  & $1 + \log  z$ & $  z^{-1}$ \\
        Itakura-Saito & $\frac{y}{z} - \log \frac{y}{z} -1 $ & $- \log z$  & $-z^{-1}$ & $z^{-2}$  \\
        beta-divergence ($\beta \in \mathbb{R}\setminus \{0,1\}$) & $\displaystyle \frac{y^{\beta}}{\beta-1} - \displaystyle \frac{\beta y z^{\beta-1}}{\beta-1} + z^{\beta}$  & $ \displaystyle \frac{z^\beta }{\beta(\beta-1)}-\frac{z}{\beta -1}+\frac{1}{\beta}$ & $ \displaystyle \frac{ z^{\beta-1}- 1 }{\beta -1}$ & $ z^{\beta -2}$ \\
    \hline
    \hline
    \end{tabular}
  \end{adjustbox}
\end{table*}

We propose to generalize the problem \eqref{eq:pr} by substituting the quadratic loss by a general Bregman divergence. A Bregman divergence $\mathcal D_\psi$ is defined from a generating function $\psi$ as follows:
\begin{equation}
\label{e:breg_D}
\mathcal D_\psi (\mathbf y\, \bm | \, \mathbf z)= \sum_{k=1}^K d_{\psi} (y(k) \bm | \,  z(k))
\end{equation}
with
\begin{equation}
\label{e:breg_d}
d_{\psi} (y \bm | \,  z)= \psi( y) - \psi( z) - \psi'(z)( y - z),
\end{equation}
where $\psi$ is a strictly-convex scalar function, continuously-differentiable on a closed convex definition domain with derivative $\psi'$, see, e.g., \cite{ban05}. We here further assume that $\psi$ is twice-differentiable with second derivative $\psi''$. $\mathcal D_\psi$ is always convex with respect to its first argument, but not necessarily with respect to its second one \cite{bauschke2001joint}.

The motivation for using Bregman divergences is two-fold. First, they encompass several divergences that are well suited for audio spectrograms such as KL or IS, as illustrated in Table~\ref{tab:bregdiv}. Second, writing those divergences under the form \eqref{e:breg_D}-\eqref{e:breg_d} will ease the derivations, as will be seen hereafter.

As $\mathcal D_\psi$ is not necessarily symmetric with respect to its input arguments, we will tackle the two following formulations of the problem:
\begin{equation}
\label{eq:bregpr_right}
    \underset{\mathbf x \in \mathbb C^L}{\text{min}}\, \overset{\rightarrow}{J}(\mathbf x):=\mathcal D_\psi(\mathbf r \,\bm |\, |\mathbf A \mathbf x|^d),
\end{equation}
\begin{equation}
\label{eq:bregpr_left}
    \underset{\mathbf x \in \mathbb C^L}{\text{min}}\, \overset{\leftarrow}{J}(\mathbf x):=\mathcal D_\psi( |\mathbf A \mathbf x|^d \,\bm |\,\mathbf r).
\end{equation}
We will refer to problems \eqref{eq:bregpr_right} and \eqref{eq:bregpr_left} as ``right PR'' and ``left PR'' respectively.

\subsection{Gradient descent and acceleration}\label{sec:3B}

Similarly to \cite{WF}, we first propose a Wirtinger gradient descent algorithm to minimize the objective functions defined in \eqref{eq:bregpr_right} and \eqref{eq:bregpr_left}.
The gradients of a general Bregman divergence with respect to its first and second arguments are given by
\begin{align}
    \nabla_\mathbf z \mathcal D _\psi (\mathbf y \,\bm | \,\mathbf z) &= \psi''(\mathbf z)\odot(\mathbf z - \mathbf y), \\
    \nabla_\mathbf y \mathcal D _\psi (\mathbf y \,\bm | \,\mathbf z) &= \psi'(\mathbf y) - \psi'(\mathbf z).
\end{align}
Using the chain rule~\cite{Magnus1985}, we obtain:
\begin{align}
    \nabla \overset{\rightarrow}{J}( \mathbf x) &=  (\nabla |\mathbf{A} \mathbf{x}  |^d)^\mathsf{H} [\psi''(|\mathbf A \mathbf x|^d) \odot
    (|\mathbf A \mathbf x|^d-\mathbf r)], \label{eq:grad1} \\
    \nabla \overset{\leftarrow}{J}( \mathbf x) &=  (\nabla |\mathbf{A} \mathbf{x}  |^d)^\mathsf{H} [\psi' (|\mathbf A \mathbf x|^d) - \psi'(\mathbf r)], \label{eq:grad2}
\end{align}
where the derivative $\psi'$ and second-derivative $\psi''$ are applied entrywise and $\nabla |\mathbf{A} \mathbf{x}  |^d$ denotes the Jacobian of the multivariate function $\mathbf{x} \to |\mathbf{A} \mathbf{x}  |^d$ (the Jacobian being the extension of the gradient for multivariate functions, we may use the same notation $\nabla$).\footnote{
Note that the gradient is not properly defined in some cases when one or more coefficients of $\mathbf A\mathbf x$ are zero-valued. We present in Appendix \ref{apdx:regrad} a detailed and rigorous treatment of this potential issue.} Using differentiation rules for element-wise matrix operations~\cite{Magnus1985}, we have:
\begin{equation}
    \label{eq:jacob}
    \nabla |\mathbf{A} \mathbf{x}|^d =
    \frac{d}{2} \,
 \text{diag}(|\mathbf{A}\mathbf{x}|^{d-2} \odot (\mathbf{A}\mathbf{x}))
    \mathbf{A}.
\end{equation}
Expressions of $\psi$, $\psi'$ and $\psi''$  for some typical Bregman divergences are given in  Table~\ref{tab:bregdiv}.

We rewrite the gradients~\eqref{eq:grad1} and \eqref{eq:grad2} in the following compact form:
\begin{equation}
    \nabla J( \mathbf x) =  (\nabla |\mathbf{A} \mathbf{x}  |^d)^\mathsf{H} \, \mathbf{g}_{\psi},
    \label{eq:nabla_J_complex}
\end{equation}
where $J$ can be either $\overset{\rightarrow}{J}$ or $\overset{\leftarrow}{J}$ and
\begin{align}
   \text{for ``right'' PR, } \mathbf{g}_{\psi} &=
   \psi''(|\mathbf A \mathbf x|^d) \odot (|\mathbf A \mathbf x|^d-\mathbf r), \\
    \text{for ``left'' PR, } \mathbf{g}_{\psi} &= \psi'(|\mathbf A \mathbf x|^d) - \psi'(\mathbf r).
\end{align}
As such and together with~\eqref{eq:jacob}, we obtain:
\begin{equation}
    \nabla J( \mathbf x) = \frac{d}{2}\, \mathbf{A}^\mathsf{H} \left[|\mathbf{A} \mathbf{x}|^{d-2} \odot (\mathbf{A} \mathbf{x}) \odot \mathbf{g}_{\psi} \right].
\end{equation}
Using a constant step-size $\mu$, our generic gradient algorithm writes:
\begin{equation}
\label{eq:grad_update_bregman}
\mathbf x_{t+1} = \mathbf x_t - \mu \nabla J(\mathbf x_t).
\end{equation}
Similarly as in FGLA \cite{FGLA}, we furthermore use a Nesterov-like acceleration scheme~\cite{NESTEROV1987} resulting in the following updates:
\begin{equation}
    \begin{aligned}
    \mathbf q_{t+1} &= \mathbf x_t - \mu \nabla J(\mathbf x_t), \\
    \mathbf x_{t+1} &= \mathbf q_{t+1} + \eta(\mathbf q_{t+1} - \mathbf q_t),
    \end{aligned}
\end{equation}
where $\eta$ is the acceleration parameter.

\noindent \textit{Remark}: When considering a quadratic loss (i.e., $\psi(z) = \frac{1}{2} z^2$), problems \eqref{eq:pr}, \eqref{eq:bregpr_right} and \eqref{eq:bregpr_left} become equivalent. In particular, when $d=1$, both gradients \eqref{eq:grad1}-\eqref{eq:grad2} write:
\begin{equation}
\label{eq:gradl2}
\nabla J(\mathbf x)= \mathbf x - \mathbf A^\mathsf H \left( \mathbf r \odot \frac{\mathbf A \mathbf x}{|\mathbf A \mathbf x|}\right).
\end{equation}
Generic gradient descent with step size equal to $1$ thus yields:
\begin{equation}
\label{eq:grad_update_gl}
\mathbf x_{t+1}=\mathbf A^\mathsf H  \left( \mathbf r \odot \frac{\mathbf A \mathbf x_t}{|\mathbf A \mathbf x_t|} \right),
\end{equation}
which is nothing but the GLA update given by alternating the projections in~\eqref{eq:glaproj}. This shows that GLA can be seen as a gradient descent applied to the PR problem~\eqref{eq:pr}. \par

\subsection{ADMM algorithm}\label{sec:3C}
In a similar fashion as in \cite{liang17}, we propose to reformulate PR with Bregman divergences as a constrained problem. We detail hereafter the left PR problem, and a similar derivation can be conducted for its right counterpart. The problem rewrites:
\begin{equation}
        \underset{\mathbf x \in \mathbb C^L, \mathbf u \in \mathbb R^K_+, \theta \in \left [0;2 \pi \right [^K}{\text{min}}\, \mathcal D_\psi(\mathbf r \,\bm|\, \mathbf u) \mbox{ subject to } (\mathbf A \mathbf x)^d = \mathbf u \odot e^{i\bm \theta},
\end{equation}
from which we obtain the augmented Lagrangian:
\begin{equation}
    \mathcal L(\mathbf x, \mathbf u, \bm \theta, \bm \lambda) = \mathcal D_\psi(\mathbf r \,\bm|\, \mathbf u) + \mathfrak R \left( \bm \lambda^{\mathsf H} ((\mathbf A\mathbf x)^d-\mathbf u \odot e^{i\bm \theta}) \right) \nonumber \\
    +\frac{\rho}{2}\left\|(\mathbf A\mathbf x)^d-\mathbf u \odot e^{i\bm \theta}\right\|_2^2 ,
\end{equation}
where $\rho$ is the penalty parameter. The first step of our ADMM algorithm consists in updating the values of $\mathbf u$ and $\bm \theta$ given $\mathbf x_t$ and $\bm \lambda_t$:
\begin{equation}
    \{\mathbf u_{t+1}, \bm \theta_{t+1}\} = \underset{\mathbf u \geq 0, \bm \theta}{\text{argmin}} \,\mathcal L(\mathbf x_t, \mathbf u, \bm \theta, \bm \lambda_t).
    \label{eq:admm_utheta}
\end{equation}
To that end, we first rewrite $\mathcal{L}$ as:
\begin{equation}
  \mathcal L(\mathbf x, \mathbf u, \bm \theta, \bm \lambda) = \mathcal D_\psi(\mathbf r \,|\, \mathbf u) \\ + \frac \rho 2 \left \|(\mathbf A \mathbf x )^d + \frac{\bm \lambda}{\rho} - \mathbf u \odot e^{i \bm \theta} \right \|_2^2
  - \frac{1}{2\rho}\left \| \bm \lambda \right \|_2^2.
  \label{eq:L}
\end{equation}
Therefore, problem~\eqref{eq:admm_utheta} can be equivalently formulated as:
\begin{equation}
    \label{eq:admm_argmin_eq}
    \{\mathbf u_{t+1}, \bm \theta_{t+1}\} = \underset{\mathbf u \geq 0, \bm \theta}{\text{argmin}} \, \mathcal D_\psi(\mathbf r \,\bm|\, \mathbf u) + \frac{\rho}{2}\|\mathbf h_t - \mathbf u\odot e^{i\bm \theta}\|_2^2,
\end{equation}
with:
\begin{equation}
    \label{eq:admm_h}
    \mathbf h_t=(\mathbf A \mathbf x_t)^d + \frac{\bm \lambda_t}{\rho}.
\end{equation}
With $\mathbf u$ fixed, the second term in \eqref{eq:admm_argmin_eq} is minimized when the phase of $\mathbf h_t$ is equal to $\bm \theta$. Thus, $\bm \theta$ is updated as follows:
\begin{equation}
  \bm \theta_{t+1}=\angle \mathbf h_t.
\end{equation}
The problem in $\mathbf u$ can then be formulated as:
\begin{equation}
  \mathbf u_{t+1} = \underset{\mathbf u \geq 0 }{\text{argmin}} \quad \mathcal D_\psi(\mathbf r \, \bm | \, \mathbf u)+\frac{\rho}{2}\||\mathbf h_t|-\mathbf u\|^2_2.
  \label{eq:ADMM_u_argmin}
  \end{equation}
As shown in Appendix~\ref{sec:apdx_prox}, the minimization problem~\eqref{eq:ADMM_u_argmin} remains unchanged when the positivity constraint on $\mathbf u$ is disregarded. The $\mathbf u$ update can therefore be written
  \begin{equation}
  \mathbf u_{t+1} = \text{prox}_{\rho^{-1}\mathcal D_\psi(\mathbf r \, \bm | \, \cdot)}(|\mathbf h_t|),
  \label{eq:ADMM_u_theta}
\end{equation}
where $\text{prox}_f$ denotes the proximal operator of a convex function $f$. The expressions of $\text{prox}_f$ for some of the divergences considered in our experiments are given in Appendix~\ref{sec:apdx_prox}.

The second step of our ADMM algorithm consists in updating the value of $\mathbf x$:
\begin{equation}
    \mathbf x_{t+1}=\underset{\mathbf x \in \mathbb C^L}{\text{argmin}} \, \mathcal L(\mathbf x, \mathbf u_{t+1}, \bm \theta_{t+1}, \bm \lambda_t).
\end{equation}
Since only the second term on the right-hand side of~\eqref{eq:L} depends on $\mathbf{x}$, this problem rewrites:
\begin{equation}
    \mathbf x_{t+1}=\underset{\mathbf x \in \mathbb C^L}{\text{argmin}} \, \left\| (\mathbf A\mathbf x)^d - \mathbf u_{t+1}\odot e^{i \bm \theta_{t+1}} + \frac{\bm \lambda_t}{\rho} \right\|_2^2,
    \label{eq:admm_prob_x_complex}
\end{equation}
which is a least-squares problem with the following closed-form solution:
\begin{equation}
    \label{eq:admm_x}
    \mathbf x_{t+1}=\mathbf A^\mathsf H\big ( \mathbf u_{t+1}\odot e^{i \bm \theta_{t+1}} -\frac{\bm \lambda_t}{\rho} \big ) ^{1/d}.
\end{equation}
The final step of our ADMM algorithm consists in updating the Lagrange multipliers $\bm \lambda$, as follows:
\begin{equation}
    \label{eq:admm_lambda}
    \bm \lambda_{t+1} = \bm \lambda_{t} + \rho (\mathbf A \mathbf x_{t+1} - \mathbf u_{t+1}\odot e^{i \bm \theta_{t+1}}).
\end{equation}
The whole ADMM procedure then consists in iteratively applying the updates given by~\eqref{eq:ADMM_u_theta},~\eqref{eq:admm_x} and~\eqref{eq:admm_lambda}.

The derivation of the updates for the left PR problem is similar, and the resulting algorithm is unchanged, except for the update of $\mathbf u$ in~\eqref{eq:ADMM_u_theta}, which becomes:
\begin{equation}\label{eq:ADMM_uR}
    \mathbf u_{t+1}= \text{prox}_{\rho^{-1}\mathcal D_\psi(\cdot\,\bm |\, \mathbf r)}(|\mathbf h_t|).
\end{equation}

\subsection{Implementation}
We have presented gradient descent and ADMM algorithms for phase retrieval in the general case. We now address some specificities of audio signal recovery from a phaseless spectrogram, i.e., when $\mathbf{A}$ is the STFT matrix and $\mathbf{x}$ is real-valued. The STFT matrix $\mathbf{A}$ and its inverse are large structured matrices that allow for fast implementations of matrix-vector products of the forms $\mathbf{A} \mathbf{x}$ and $\mathbf{A}^{\mathsf{H}} \mathbf{y}$ based on the fast Fourier transform~\cite{Brigham1967,smith2011spectral}. In that setting, one handles time-frequency matrices of size $M \times N$, where $M$ is the number of frequency channels and $N$ the number of time frames, rather than vectors of size $K=MN$. As such, we provide in Algorithms~\ref{al:bregman_gradient} and~\ref{al:bregman_admm} the pseudo-code for practical implementation of our accelerated gradient and ADMM algorithms, respectively, in the time-frequency audio recovery setting.

\begin{algorithm}[h]
	\caption{Accelerated gradient descent for PR with the Bregman divergence.}
	\label{al:bregman_gradient}
			\textbf{Inputs}: Measurements $\mathbf{R} \in \mathbb{R}_+^{M \times N}$, initial phase $\boldsymbol{\phi}_0 \in [0,2\pi[_+^{M \times N}$, step size $\mu$ and acceleration parameter $\eta$. \\

            \textbf{Initialization}: \\
            $\mathbf{X} = \mathbf{R}^{1/d} \odot e^{i\boldsymbol{\phi}_0}$\\
            $\mathbf{x} = \text{iSTFT}(\mathbf{X})$ \\
            $\mathbf q_{\text{old}} = 0$

			\While{stopping criteria not reached}{

			$\mathbf X = \text{STFT}(\mathbf{x})$ \\
  \uIf{PR left}{
    $\mathbf G_{\psi} = \psi' (|\mathbf{X}|^d) - \psi' (\mathbf{R})$
  }
  \uElseIf{PR right}{
    $\mathbf G_{\psi} =  \psi'' (|\mathbf{X}|^d) \odot  ( |\mathbf{X}|^d -\mathbf{R})$
  }
            $\mathbf q = \mathbf x - \mu \frac{d}{2} \text{iSTFT}(\mathbf X \odot |\mathbf X|^{d-2} \odot \mathbf G_{\psi} )  $ \\
            $\mathbf x = \mathbf q + \eta(\mathbf q - \mathbf q_{\text{old}})$ \\
            $\mathbf q_{\text{old}} = \mathbf q$
			}

			\textbf{Output}: $\mathbf{x}$

\end{algorithm}

\begin{algorithm}[h]
	\caption{ADMM for PR with the Bregman divergence.}
	\label{al:bregman_admm}
			\textbf{Inputs}: Measurements $\mathbf{R} \in \mathbb{R}_+^{M \times N}$, initial phase $\boldsymbol{\phi}_0 \in [0,2\pi[_+^{M \times N}$ and augmentation parameter $\rho$. \\

            \textbf{Initialization}: \\
            $\mathbf{X} = \mathbf{R}^{1/d} \odot e^{i\boldsymbol{\phi}_0}$\\
            $\mathbf{x} = \text{iSTFT}(\mathbf{X})$ \\
            $\boldsymbol{\Lambda} = 0$

			\While{stopping criteria not reached}{

			$\mathbf X = \text{STFT}(\mathbf{x})$ \\
			$\mathbf H = \mathbf{X}^d + \frac{1}{\rho} \boldsymbol{\Lambda}$ \\
			$\boldsymbol{\Theta} = \angle \mathbf H$\\

  \uIf{PR left}{
    $\mathbf U = \text{prox}_{\rho^{-1}\mathcal D_\psi(\cdot\,\bm |\, \mathbf r)}(|\mathbf H|)$
  }
  \uElseIf{PR right}{
    $\mathbf U = \text{prox}_{\rho^{-1}\mathcal D_\psi(\mathbf r\,\bm |\, \cdot)}(|\mathbf H|)$
  }
			$\mathbf{x} = \text{iSTFT}( (\mathbf{U} \odot e^{i\boldsymbol{\Theta}} - \frac{1}{\rho} \boldsymbol{\Lambda})^{1/d})$ \\
			$\boldsymbol{\Lambda} = \boldsymbol{\Lambda} + \rho (\text{STFT}(\mathbf{x})-\mathbf{U} \odot e^{i\boldsymbol{\Theta}})$\\
			}

			\textbf{Output}: $\mathbf{x}$

\end{algorithm}

For generality, we assumed $\mathbf{x} \in \mathbb{C}^L$ in the previous sections. However, audio signals are real-valued and this deserves some comments. 
As shown in Appendix \ref{sec:apdx_real}, the estimates $\mathbf x_t$ remain real-valued under the following conditions. In a nutshell, a signal is real-valued if and only if its STFT $\mathbf X \in \mathbb C^{M\times N}$ is frequency-Hermitian, that is:
\begin{equation}
    X (m,n) = X(M-m,n)^*.
\end{equation}
When $\mathbf{R}$ is the spectrogram of a real-valued signal and when Algorithms~\ref{al:bregman_gradient} and~\ref{al:bregman_admm} are initialized with a frequency-Hermitian matrix $\mathbf{X}$, 
all the time-frequency matrices involved in the updates remain frequency-Hermitian (because operations only involve sum and element-wise product with frequency-Hermitian matrices).
This in turn ensures that the variable $\mathbf{x}$ remains real-valued. As such, the STFT and inverse STFT (iSTFT) operations in Algorithms~\ref{al:bregman_gradient} and~\ref{al:bregman_admm} need only return/process the first $\floor*{\frac{M}{2}}+1$ frequency channels (usually termed ``positive frequencies"), as customary with real-valued signals~\cite{Sondergaard2011}.

More rigorously, we may also re-derive our gradient and ADMM algorithms for $\mathbf{x} \in \mathbb{R}^L$, using real-valued differentiation instead of Wirtinger gradients (and involving the real and imaginary parts of $\mathbf{A}$ in the objective function). This is addressed in Appendix~\ref{sec:apdx_real} which shows that we indeed obtain the same algorithms.

\section{Experiments}
\label{sec:exp}

In this section, we conduct experiments on  PR tasks. We first assess the potential of the proposed algorithms for recovering signals from exact (i.e., non-modified) spectrograms. Then, we consider a PR task from modified spectrograms, as often encountered in audio applications. In the spirit or reproducible research, the code related to those experiments is available online.\footnote{\url{https://github.com/phvial/PRBregDiv}} We also provide audio examples of reconstructed signals.\footnote{\url{https://magronp.github.io/demos/jstsp21.html}}

\subsection{Experimental setup}
\label{sec:exp_protocol}

\subsubsection{Data}
\label{sec:exp_protocol_data}

As acoustic material, we use two corpora in our experiments. The first one, referred to as ``speech'', is composed of $100$ utterances taken randomly from the TIMIT database~\cite{Garofolo1993}. The second one, referred to as ``music'', comprises $100$ snippets from the Free Music Archive dataset \cite{fma_dataset}.
Signals from the ``speech'' corpus are $16$-bits WAV files and signals from the ``music'' corpus are MP3 files encoded at $256$ kbps. All audio excerpts are single-channel, sampled at $22,050$ Hz and cropped to be $2$ seconds long.  The STFT is computed with a $1024$ samples-long ($46$ ms) self-dual sine window~\cite{smith2011spectral} (leading to an effective number of $513$ frequency bins) and $50$~$\%$ overlap. We used the \texttt{librosa} Python package~\cite{mcfee2015librosa}.

\subsubsection{Methods}
\label{sec:exp_protocol_methods}

PR is conducted using the algorithms presented in Section~\ref{sec:pr_bregman} under different settings as described next.

\paragraph{Proposed gradient descent algorithm}
We experimented the accelerated gradient algorithm described in Alg.~\ref{al:bregman_gradient} in the following settings:

\begin{itemize}
    \item KL ($\beta=1$) for the ``right'' and ``left'' problems with $d \in \{1,2\}$,
    \item $\beta=0.5$ for the ``right'' and ``left'' problems and with $d \in \{1,2\}$,
\item IS ($\beta=0$) for the ``right'' problem with $d=2$,
    \item quadratic loss ($\beta=2$) with $d \in \{1,2\}$ (in that case the ``right'' and ``left'' problems are equivalent).
\end{itemize}
The ``right'' problems with KL, $d=1$ on the one hand, and IS, $d=2$ on the other hand, correspond to standard designs in NMF \cite{smaragdis14,Fevotte2018a}. The trade-off value $\beta=0.5$ with either $d=1$ or $2$ has also been advocated in various papers, e.g.,~\cite{vinc10}.

The algorithms are used with constant step-size $\mu$ and acceleration parameter $\eta=0.99$ (like in \cite{FGLA}). The step-size is empirically set to the largest negative power of $10$ enabling convergence for each loss and value of $d$ in the setting of the experiments reported in Sections~\ref{sec:exp_rec_spectro} and~\ref{sec:exp_speech}. A summary of the parameter configurations and choice of loss functions is given in Table \ref{tab:recap_alg}.

\begin{table*}[!htbp]
    \centering
    \caption{Summary of setups considered in the experiments with their parameters (loss function, exponent $d$, type of algorithm and hyperparameter). Each setup is described by a code that follows this format: \textit{algorithm$\cdot$loss$\cdot$direction-d}.}
    \begin{adjustbox}{angle=90}
    \begin{tabular}{|c|c|c|c|c|c|c|} \hline
        Algorithm & Gradient descent & Gradient descent & Gradient descent & Gradient descent & Gradient descent & Gradient descent   \\
        Loss & $\beta$-div. ($\beta=0.5$) & $\beta$-div. ($\beta=0.5$)& Kullback-Leibler & Kullback-Leibler& Quadratic & Itakura-Saito \\
        Direction & right  & left& right & left & N/A & right\\
        $d$ & $1$ & $1$ & $1$& $1$& $1$ & $2$\\
        Hyperparameters & $\mu=10^{-1}$ & $\mu=10^{-6}$ & $\mu=10^{-4}$& $\mu=10^{-2}$&$\mu=10^{-1}$ &$\mu=10^{-7}$\\
        Associated code & G$\cdot$05$\cdot$R1   & G$\cdot$05$\cdot$L1 & G$\cdot$KL$\cdot$R1 & G$\cdot$KL$\cdot$L1& G$\cdot$QD$\cdot$1& G$\cdot$IS$\cdot$R2 \\ \hline\hline

        Algorithm & Gradient descent & Gradient descent & Gradient descent & Gradient descent & Gradient descent~\cite{WF} & ADMM   \\
        Loss & $\beta$-div. ($\beta=0.5$)& $\beta$-div. ($\beta=0.5$)& Kullback-Leibler& Kullback-Leibler& Quadratic& Itakura-Saito\\
        Direction & right  & left & right & left & N/A& left\\
        $d$ & $2$ & $2$ & $2$ & $2$ & $2$ & $1$\\
        Hyperparameters & $\mu=10^{-3}$ &  $\mu=10^{-6}$ &$\mu=10^{-1}$ &$\mu=10^{-3}$ &$\mu=10^{-5}$ & $\rho=10^{-1}$\\
        Associated code & G$\cdot$05$\cdot$R2   & G$\cdot$05$\cdot$L2& G$\cdot$KL$\cdot$R2& G$\cdot$KL$\cdot$L2& G$\cdot$QD$\cdot$2& A$\cdot$IS$\cdot$L1\\ \hline \hline

        Algorithm & ADMM & ADMM~\cite{liang17} & Griffin-Lim~\cite{GLA} & Fast Griffin-Lim~\cite{FGLA} & GLADMM~\cite{Masuyama2019} & Initialisation   \\
        Loss & Kullback-Leibler & Quadratic & (Quadratic) & (Quadratic) & (Indicator function) & N/A\\
        Direction &  left  & N/A & N/A & N/A& N/A& N/A\\
        $d$ & $1$ & $1$ & $1$ & $1$ & $1$ & N/A\\
        Hyperparameter &  $\rho=10^{-1}$& $\rho=10^{-1}$& N/A & N/A & N/A & N/A\\
        Associated code & A$\cdot$KL$\cdot$L1   & A$\cdot$QD$\cdot$1 & GLA & FGLA& GLADMM& INIT  \\ \hline
    \end{tabular}
\end{adjustbox}
\label{tab:recap_alg}
\end{table*}

\paragraph{Proposed ADMM algorithm}
Applicability of ADMM is more limited than with gradient descent because it requires the expression of the proximal operators \eqref{eq:ADMM_u_theta} and \eqref{eq:ADMM_uR}. We here consider the quadratic loss and ``left" KL and IS problems. We set $d=1$ and $\rho =1$, which corresponds to the setting used by Liang et al. \cite{liang17} for the quadratic loss (which thus falls as a special case of our setting).

\paragraph{Other baselines and parameters}

The previous algorithms are compared with the following other baselines: GLA, FGLA and GLADMM, presented in Section~\ref{sec:related_work} and which use $d=1$.

All the algorithms (baseline and contributed) are run for $2500$ iterations (which ensures that convergence is observed for all algorithms) and initialized with the same uniform random phase (a single realization was used for each excerpt).

\subsubsection{Evaluation metrics}
The reconstruction quality is evaluated in the time-frequency domain with the standard spectral convergence (SC) metric, which is defined in dB as:
\begin{equation}
    \text{SC}(\mathbf r, \mathbf x)= - 20 \log_{10} \left( \frac{\|\mathbf r^{1/d} - |\mathbf A \mathbf x|\|_2}{\|\mathbf r\|_2} \right).
\end{equation}
Additionally, for the ``speech'' corpus, we consider the short-term objective intelligibility (STOI) measure \cite{STOI}, which is computed with the \texttt{pystoi} library \cite{pystoi}. This score is obtained by first decomposing the clean and processed speech signals through a DFT-like filterbank, and then computing the correlation between the resulting representations' time envelopes. It has been shown to correlate well with subjective intelligibility measurements of speech, whether in clean or noisy conditions and at various subjective intelligibility ranges~\cite{STOI}. Consequently, it has been used in several PR-related papers such as~\cite{Masuyama2019deepGLA,Masuyama2019}.

Let us note that alternative evaluation metrics exist, such as PESQ~\cite{Rix2001} or PEMO-Q~\cite{Huber2006}, which are tailored for perceptually assessing speech quality. We also computed those, and the obtained results were overall consistent with the STOI measure, up to some minor differences. Besides, it has been shown that these measures are strongly correlated with STOI, notably in PR-related tasks~\cite{Mayer2017}. For these reasons, and for brevity, we did not include these here.

To summarize, SC is directly related to the PR quadratic loss problem~\eqref{eq:pr}, formulated in the time-frequency domain. On the other hand, the perceptual STOI is more related to the applicative needs. In both cases, the higher value, the better performance.

\subsection{PR from exact spectrograms}
\label{sec:exp_rec_spectro}

First, we consider a PR task conducted on exact spectrograms. In this setting, measurements are directly obtained from the ground truth signals $\mathbf x^\star$, such that $\mathbf r = |\mathbf A \mathbf x^\star|^d$. These measurements $\mathbf r$ are then fed as inputs to the algorithms described in~\ref{sec:exp_protocol_methods}.

\begin{figure}[h!]
    \centering
    \begin{subfigure}{1\linewidth}
    \centering
    \includegraphics[scale=0.23]{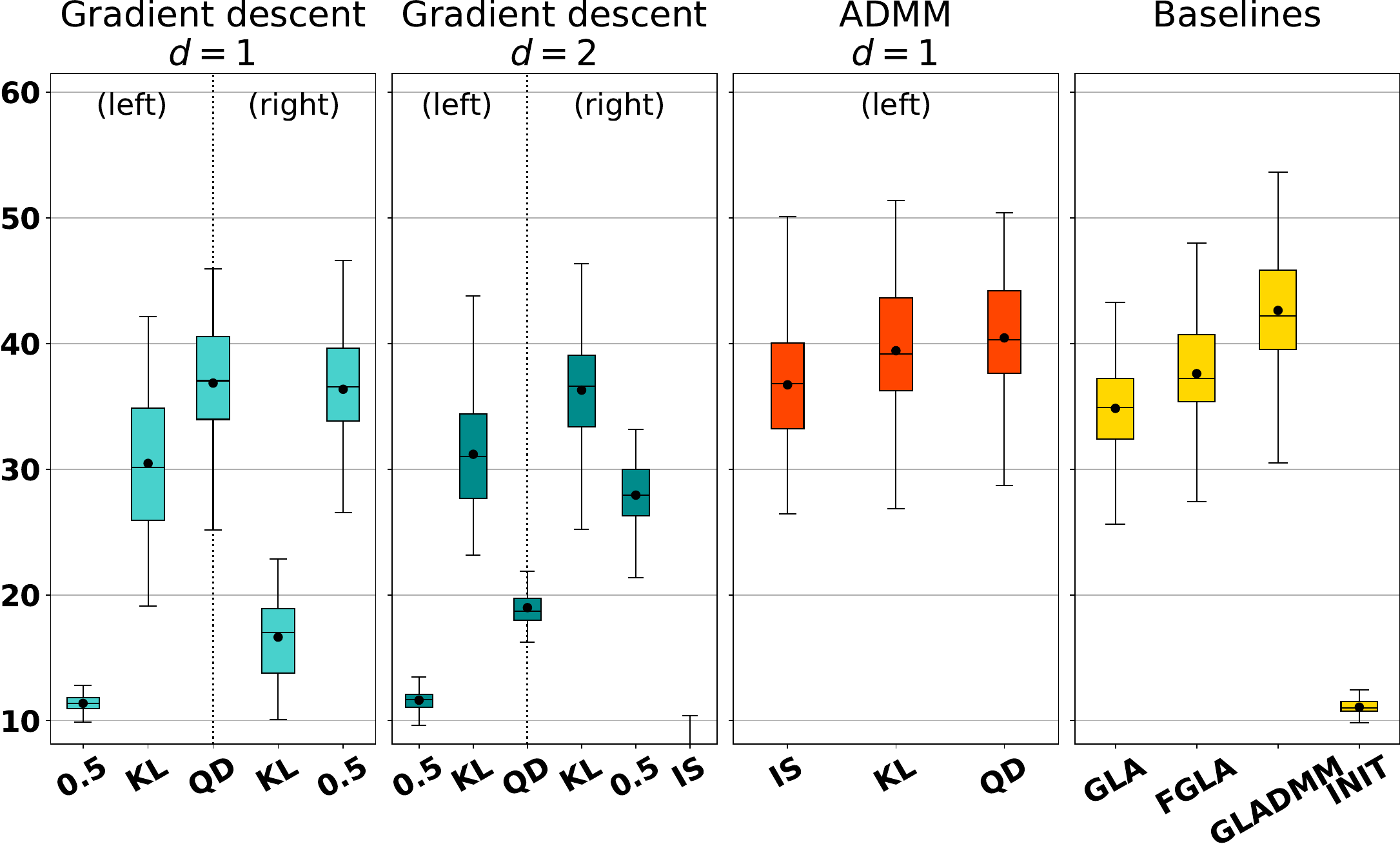}
    \vspace{-0.3cm}
    \subcaption{Spectral convergence}
    \end{subfigure}
    \vspace{0.6cm}
    \newline
    \begin{subfigure}{1\linewidth}
    \centering
    \includegraphics[scale=0.23]{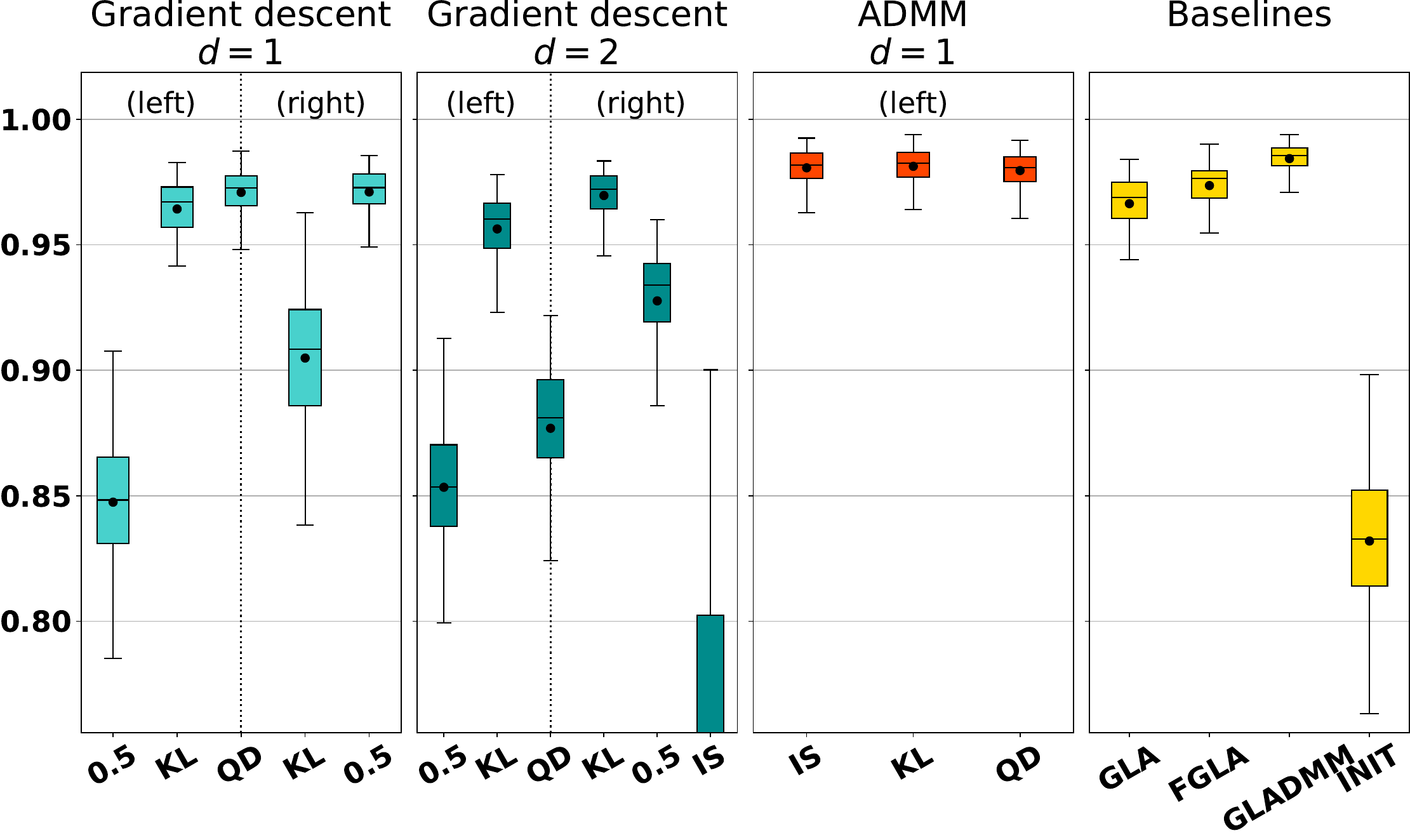}
    \vspace{-0.3cm}
    \subcaption{STOI}
    \end{subfigure}
    \vspace{0.3cm}
    \caption{Performance of PR from exact spectrograms for the ``speech'' corpus, measured with the SC (top) and STOI (bottom). Higher values correspond to a better performance. Turquoise, orange and yellow respectively denote gradient descent algorithms, ADMM algorithms and GLA-like algorithms. The boxes indicate the two middle quartiles among the ten excerpts, the middle bar is for the median, the dot for the mean, and the whiskers denote the extremal values.}
    \label{fig:reco_timit}
\end{figure}

\begin{figure}[t]
    \centering
    \begin{subfigure}{1\linewidth}
    \centering
    \includegraphics[scale=0.23]{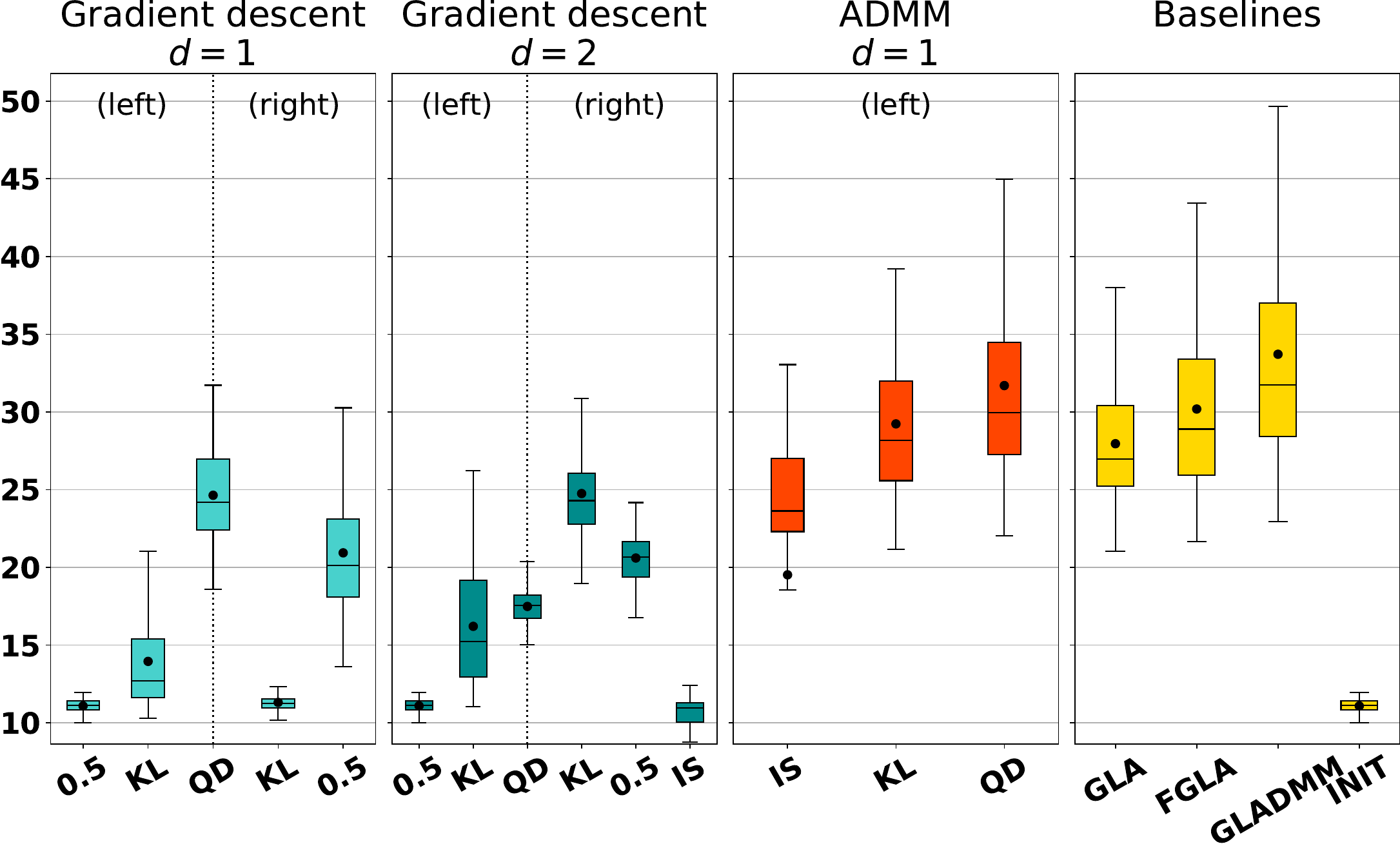}
    \end{subfigure}
    \newline \vspace{0.25cm}
    \caption{Performance of PR from exact spectrograms for the ``music'' corpus measured with the SC.}
    \label{fig:reco_music}
\end{figure}

The results on the ``speech" and ``music" corpora are presented in Figures~\ref{fig:reco_timit} and~\ref{fig:reco_music} respectively, from which overall similar conclusions can be drawn.

The best performances in terms of SC are achieved by GLADMM and other ADMM algorithms, which are closely followed by algorithms optimizing the quadratic loss with $d=1$. Note however that the advantage of quadratic loss-based algorithms against competing methods is less significant in terms of STOI. As recalled above, SC is directly related to the PR problem with quadratic loss~\eqref{eq:pr} and consequently favors algorithms that directly tackle this problem.

A performance similar to that of quadratic loss-based algorithms is reached by some of the proposed alternative methods, such as the ADMM algorithms A$\cdot$IS$\cdot$L1 and A$\cdot$KL$\cdot$L1 and the gradient descent algorithms GD$\cdot$05$\cdot$R1, GD$\cdot$KL$\cdot$R2 and GD$\cdot$KL$\cdot$L2, in terms of SC and STOI (note that for the latter, the best performing methods exhibit a lower variance than the others). This outlines the potential of using alternative divergences to the quadratic loss.

Besides, we observe that the performance of these methods depend on a variety of factors. For instance, the difference between the performance reached by GD$\cdot$KL$\cdot$R2 and GD$\cdot$KL$\cdot$R1, or between GD$\cdot$QD$\cdot$1 and GD$\cdot$QD$\cdot$2 (for both metrics and corpora) outlines the impact of $d$ on the reconstruction quality. Likewise, considering a ``left" rather than a ``right" PR problem may yield very different results (see for instance the two corresponding gradient algorithms with $\beta=0.5$ and $d=1$).

Finally, for a given problem, the impact of the optimization strategy (i.e., ADMM vs. gradient descent) depends on the nature of the signals. For the ``speech" corpus, ADMM algorithms (for KL and the quadratic loss) perform mildly better than their gradient algorithms respective counterparts. However, for the ``music" corpus, A$\cdot$KL$\cdot$L1 significantly outperforms GD$\cdot$KL$\cdot$L1 in terms of SC.

To summarize, when retrieving a signal from an exact spectrogram, GLADMM and quadratic-minimizing algorithms (with $d=1$) seem to perform best. Some alternative methods yield competitive results, but require to carefully adapt the setup (power $d$, loss $\beta$, ``right" or ``left" formulation) and optimization strategy (ADMM vs. gradient descent) to the problem, as well as considering the nature of the signals (speech or music). Note than when the data $\ve{r}$ is an exact spectrogram (i.e., $\ve{r} = |\ve{A} \ve{x}^\star|^d$), the loss functions~\eqref{eq:bregpr_right} and \eqref{eq:bregpr_left} share the same minimum value 0 and global solution $\ve{x}^\star$ (up to ambiguities) for all $\psi$. This may explain why the somehow easier-to-optimize quadratic loss performs well in this scenario. However this result is to be contrasted when using modified spectrograms, as shown next.

\subsection{PR from modified spectrograms}
\label{sec:exp_speech}

We now consider a PR task from modified spectrograms. In audio restoration applications such as source separation \cite{Vincent2018}, audio inpainting~\cite{Adler2012} or time-stretching~\cite{Driedger2016}, the spectrogram that results from diverse operations does not necessarily correspond to the magnitude of the STFT of a signal. We propose to simulate this situation by modifying the spectrograms as in~\cite{Masuyama2019}. We add synthetic Gaussian white noise in the time domain to each excerpt in the ``speech" corpus. For each signal, the noise variance is chosen so that the input signal-to-noise ratio (SNR) takes the following values: $10$, $0$, $-10$, and $-20$ dB. We then apply an oracle Wiener filter~\cite{Liutkus2015} to the mixture in the STFT domain: this yields a restored (even though inconsistent, \textit{cf}. Section~\ref{sec:2A}) magnitude spectrogram $\mathbf r$ which corresponds to realistic applications~\cite{Masuyama2019}. To further recover a time-domain signal, we apply the considered PR algorithms to this modified spectrogram.

\begin{figure}
    \centering
    \begin{subfigure}{1\linewidth}
    \centering
    \includegraphics[scale=0.23]{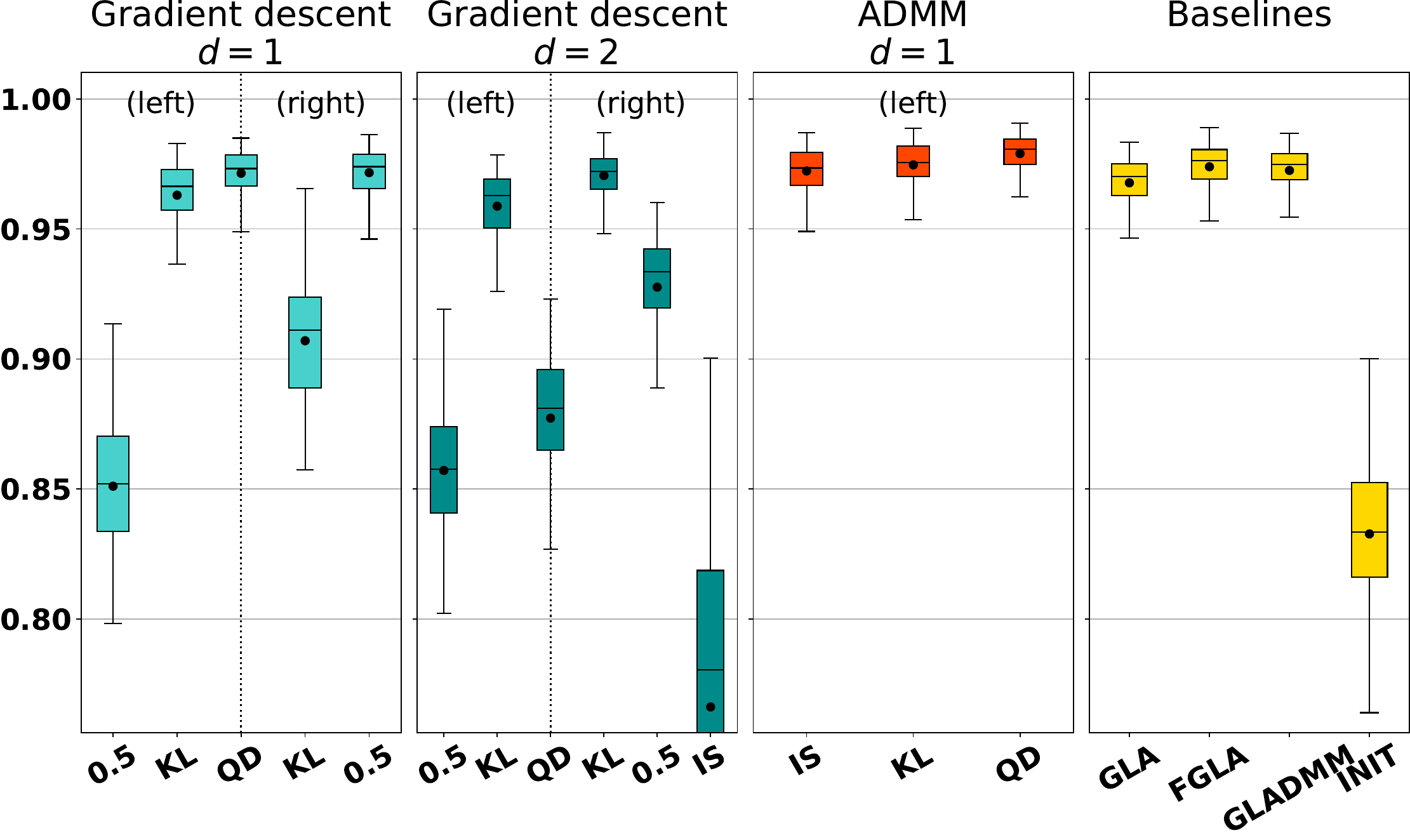}
    \vspace{-0.4cm}
    \subcaption{$+10$ dB}
    \end{subfigure}
    \vspace{0.6cm}
    \newline
    \begin{subfigure}{1\linewidth}
    \centering
    \includegraphics[scale=0.23]{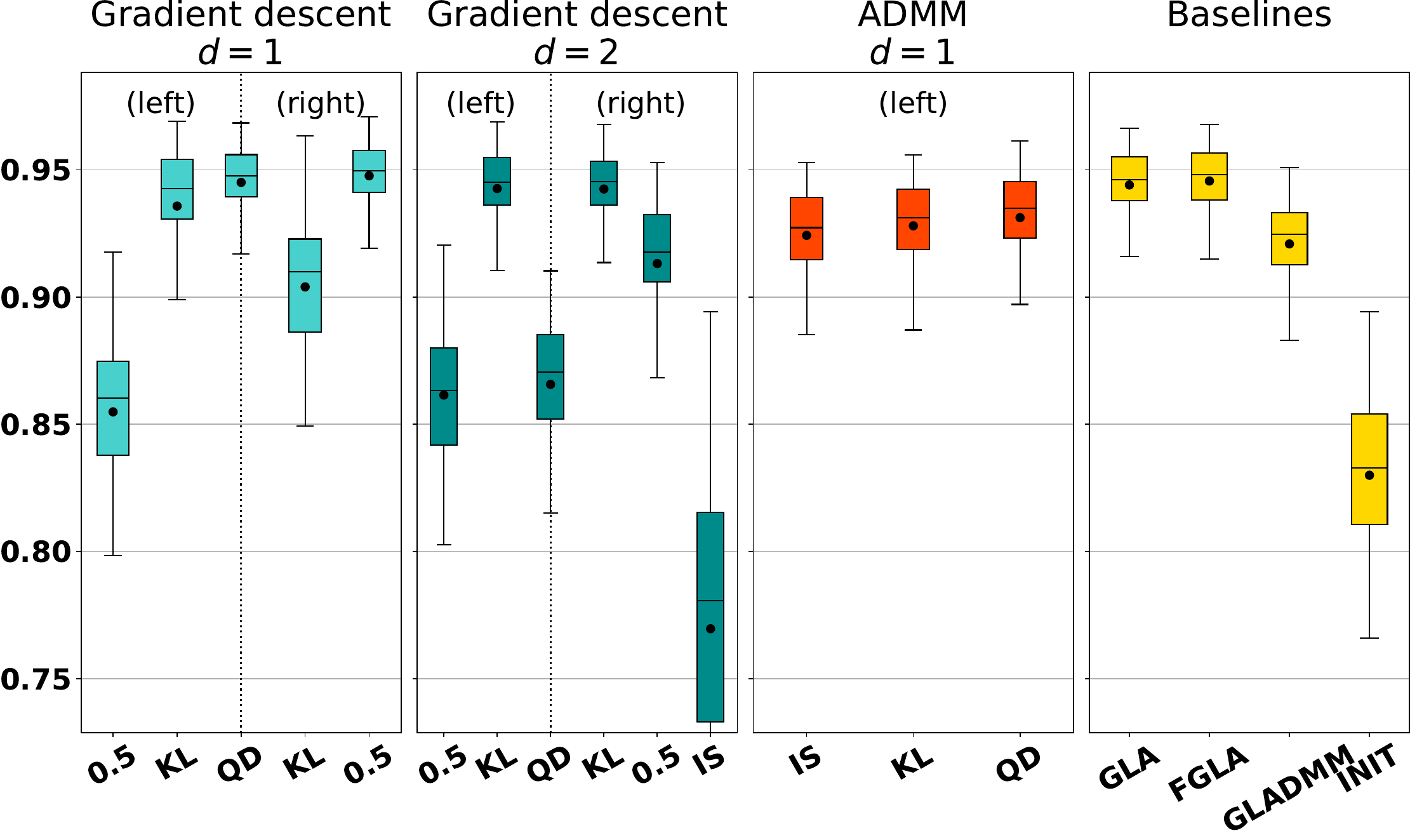}
    \vspace{-0.4cm}
    \subcaption{$0$ dB}
    \end{subfigure}
    \vspace{0.6cm}
    \newline
    \begin{subfigure}{1\linewidth}
    \centering
    \includegraphics[scale=0.23]{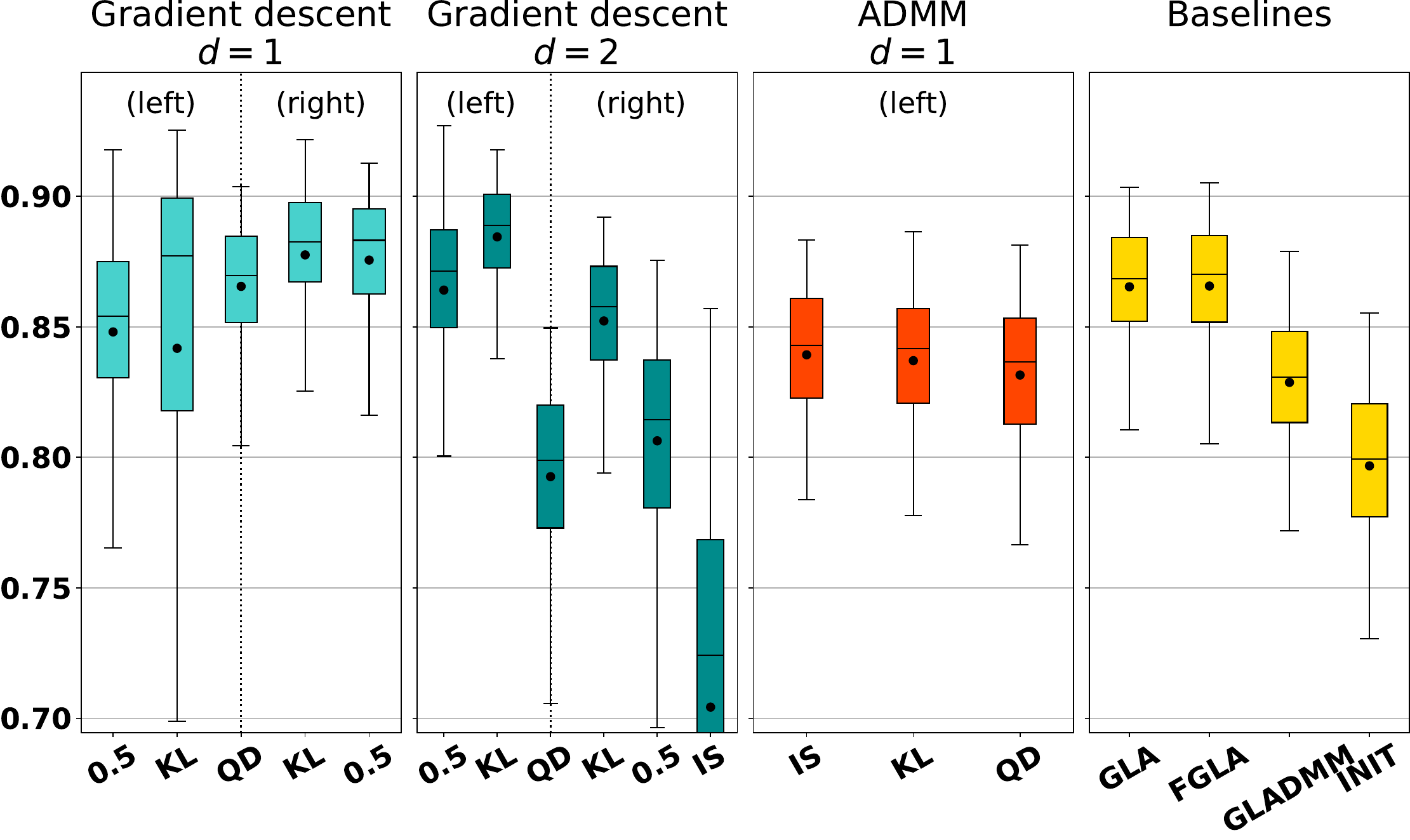}
    \vspace{-0.4cm}
    \subcaption{$-10$ dB}
    \end{subfigure}
    \vspace{0.6cm}
    \newline
    \begin{subfigure}{1\linewidth}
    \centering
    \includegraphics[scale=0.23]{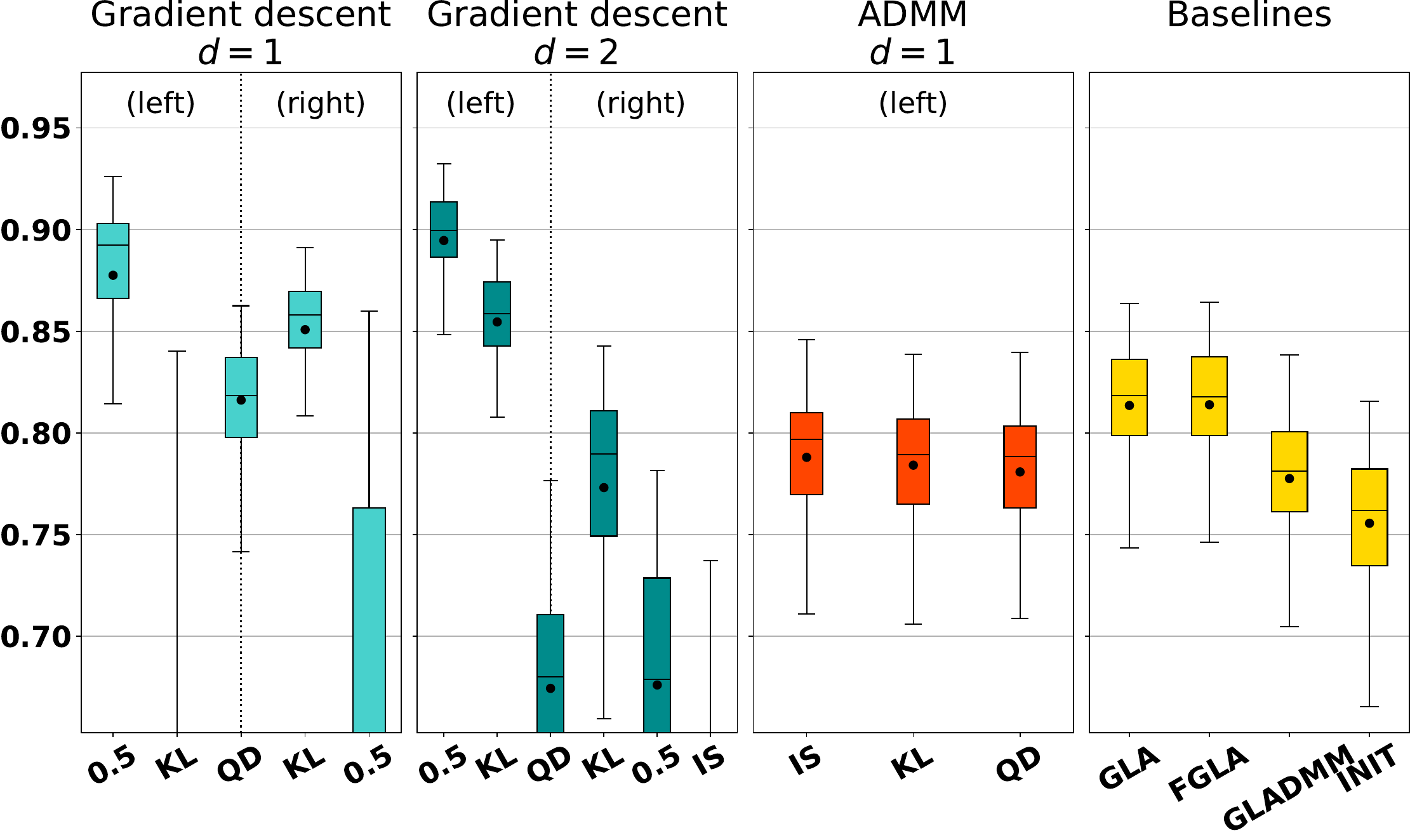}
    \vspace{-0.4cm}
    \subcaption{$-20$ dB}
    \end{subfigure}
    \caption{STOI for PR from modified speech spectrograms at various input SNRs.}
    \label{fig:deno_stoi}
\end{figure}

The results in terms of STOI are presented in Figure~\ref{fig:deno_stoi}. Note that we do not report the SC, since it is mostly impacted by the spectrogram deformation procedure, not by the subsequent PR task. In that experiment, we observed some convergence problems at low input SNRs for several algorithms and signals: in these few cases, the gradient step size (which we recall was tuned using exact spectrogram data) was reduced by a factor $1/10$.

At high input SNR ($0$ to $10$ dB), we observe a similar trend than in the previous experiment: GLADMM and quadratic loss-based algorithms (with $d=1$) enable better reconstruction in terms of STOI than other categories of algorithms. This confirms that such algorithms are appropriate for addressing the PR problem when the spectrograms are either exact or slightly degraded.

However, we observe a different trend at lower input SNRs, where some algorithms based on alternative losses exhibit more robustness to the spectrogram degradation caused by the input noise. For instance, while ADMM algorithms overall perform best at $10$ dB input SNR, they are outperformed by alternative algorithms such as GD$\cdot$KL$\cdot$L2 at lower input SNRs (from $0$ to $-20$ dB). Similarly and contrary to the case of high input SNRs, GLADMM is outperformed by other GL-based or ADMM algorithms. Interestingly, GD$\cdot$05$\cdot$L1 and GD$\cdot$KL$\cdot$R1 exhibit the most robust behavior among gradient algorithms with $d=1$: while they perform worst at $10$ dB input SNR, they actually achieve the best performances at $-20$ dB input SNR. On the other hand, the performance of several algorithms, such as GD$\cdot$KL$\cdot$R2, significantly drops when more noise is added, while they perform relatively well at high input SNRs. Finally, even though the best performance at very low input SNR is achieved by GD$\cdot$05$\cdot$L2, GD$\cdot$KL$\cdot$L2 might still be a good candidate for the task at hand: indeed, at input SNRs from $10$ to $-10$ dB, it outperforms its counterpart using $\beta=0.5$, and thus exhibits a more stable performance regarding the level of input noise.

Overall, the usefulness of PR with general Bregman divergences is revealed when the spectrograms are highly corrupted (that is, when they are retrieved using a Wiener filter from very noisy observations), as quadratic loss-based algorithms are outperformed by alternative loss-based algorithms in such a scenario. This might be explained by the ability of such divergences to better model and account for the nature of this destructive noise.

\section{Conclusion}
\label{sec:conclusion}

We have addressed the problem of PR when the quadratic loss is replaced by general Bregman divergences, a family of discrepancy measures with special cases that are well-suited for audio applications. We derived a gradient algorithm and an ADMM scheme for solving this problem and implemented them in the context of audio signal recovery. We evaluated the performance of these algorithms for PR from exact and modified spectrograms.
We experimentally observed that when performing PR from exact or slightly degraded spectrograms, traditional algorithms based on the quadratic loss perform best. However, in the presence of high level of noise, these are outperformed by algorithms based on alternative losses. This highlights the potential of PR with the Bregman divergence for audio signal recovery from spectrograms under very noisy conditions. However it is difficult to recommend a specific alternative divergence at this stage. The choice is dependent on the amount of noise and possibly on the nature of the data itself (e.g., speech vs music). Gradient algorithms are very convenient because they can be applied to any setting, however finding efficient step sizes in every setting was challenging and this issue deserves more attention. In that respect, our ADMM algorithms appeared more stable with respect to the level of noise and to the nature of the data but their applicability is more limited as they depend on the availability of specific proximal operators for each setting.

In future work, we intend to further improve the proposed gradient descent algorithms, notably by leveraging more refined initialization schemes, and to explore other optimization strategies such as majorization-minimization. It would be also useful to conduct subjective listening tests to fully assess the potential of using Bregman divergences for a phase retrieval task. Finally, we intend to tackle PR with non-quadratic measures of fit in frameworks where some additional phase information is available, such as in speech enhancement and source separation applications.

\appendix
\section{Short-time Fourier transform}
\label{sec:apdx_stft}

Given a signal $\mathbf x \in \mathbb C^L$ and an {\em analysis window} $\mathbf w \in \mathbb R^T$ such that $T<L$, the discrete short-time Fourier transform (STFT) is a linear application ${\cal A}_w$ defined by
\begin{equation}
    \label{eq:stft}
    [\mathcal{A}_w \mathbf x](m,n) := \sum_{t=0}^{T-1} x(t+nH)  w(t)e^{-i 2\pi  \frac{m}{M}t},
\end{equation}
where
\begin{itemize}
    \item $n=0,\ldots, N-1$ indexes time frames,
    \item $m=0,\ldots, M-1$ indexes frequency bins,
    \item $H$ is a ``hop" size.
\end{itemize}
The STFT essentially chops the signal $\ve{x}$ into windowed segments of size $T$ and applies a DFT of size $M$ to each segment. $H$ controls the overlap between segments. $H$ and $M$ are user-defined. Generally, $H \le T$. When $T$ is even, $H=T/2$ corresponds to a 50\% overlap between segments, which is a common choice. Generally, $M \ge T$ (more frequencies than samples). A common choice is $M=T$, which corresponds to using a standard ``square" DFT. The value of $N$ is determined by the length of the signal $L$, the length of the window $T$ and the hop-size $H$. Common practice consists in zero-padding the signal $\ve{x}$ with $T-H$ zeroes at the beginning and as many zeroes as needed at the end so that $L = (T-H) + NH = T + (N-1)H$. This is in particular needed to have perfect reconstruction at the borders when defining an inverse-operator. We here assume that the signal $\ve{x}$ (of length $L$) has undergone such zero-padding at its borders.

Given a time-frequency matrix $\ve{C} \in \mathbb C^{M \times N}$ and a {\em synthesis window} $\mathbf v \in \mathbb R^T$, an inverse-STFT can be defined through the linear application $\mathcal{S}_{v}$ defined by
\begin{equation}
\label{eq:istft}
    [\mathcal{S}_{v}\mathbf C] (\ell):=
    \frac{1}{M} \sum_{n=0}^{N-1}\sum_{m=0}^{M-1}  C(m,n)  v(\ell - nH) e^{i 2\pi  \frac{m}{M} (\ell-nH)  },
\end{equation}
where $\ell = 0,\ldots, L-1$. We use the convention that $w(t) = v(t) = 0$ whenever $t \notin [0, T-1]$. The inverse-STFT essentially applies an inverse DFT to each time-frame of $\ve{C}$ and overlap-adds the resulting temporal signals. The windows $\ve{w}$ and $\ve{v}$ are said to be \emph{dual} whenever they satisfy the following condition:
\begin{equation}
    \forall \ell\, , \quad \sum _{n=0}^{N-1}  w(\ell - nH)v(\ell - nH) = 1.
\end{equation}
In this case (and when $M \ge T$), perfect reconstruction is achieved~\cite{smith2011spectral}, i.e.,
\begin{equation}
    \mathcal S_{v} \mathcal A_{w} \mathbf x = \mathbf x.
\end{equation}
The STFT can alternatively be written as a Gabor frame. Indeed,~\eqref{eq:stft} can be written as the output of inner products between $\ve{x}$ and Gabor atoms $\bm \gamma_{mn} \in \mathbb C^L$ defined as pure windowed complex exponential, such that
\begin{equation}
    \gamma_{mn}(\ell) = w(\ell-nH) e^{i 2\pi \frac{m}{M}(\ell-nH)}.
\end{equation}
Ignoring the time-frequency ordering and collecting the Gabor atoms into the columns of an $L \times MN $ matrix $\bm \Gamma_w$, the STFT can equivalently be obtained by $\bm \Gamma_w^\mathsf{H} \ve{x}$ (and as such $\bm \Gamma_w^\mathsf H$ is equal to the matrix $\ve{A}$ used in the main body of the paper). Under general conditions \cite{grochenig2013foundations}, the matrix $\bm \Gamma_w$ defines a frame in the sense that there exists positive constants $\alpha_1$ and $\alpha_2$ such that
\begin{equation}
    \alpha_1 \|\mathbf x\|^2_2 \leq \|\bm \Gamma_w^\mathsf{H} \mathbf x\|^2_2 \leq \alpha_2  \|\mathbf x\|^2_2.
\end{equation}
Similarly, the synthesis operator $\mathcal{S}_v$ can be expressed as
\begin{equation}
    \mathcal{S}_v \ve{C} = \bm \Gamma_v \ve{c}
\end{equation}
where $\ve{c}$ is a vectorized version of $\ve{C}$. As such, the windows $\mathbf w$ and $\mathbf v$ are dual if and only if $\bm \Gamma_v \bm \Gamma_w^\mathsf H \ve{x} = \ve{x}$. When the same window can be used for analysis and synthesis with perfect reconstruction (an example being the sine window \cite{smith2011spectral}), then it can be shown that $\alpha_1=\alpha_2=1$ and $\bm \Gamma_w^\mathsf H$ defines a so-called {\em Parseval frame}. This last assumption holds everywhere in the main body of the paper (i.e., $\ve{A}^\mathsf H \ve{A} = \ve{I}_L$).

\section{Wirtinger formalism}
\label{sec:apdx_wirtinger}

A function $f$, which can be either complex- or real-valued, of a complex variable $x=x_r+ i x_i$ can be seen as a function of $(x_r,x_i)$. The Wirtinger derivatives~\cite{wirtinger1927formalen, bouboulis2010extension, Bouboulis2010} are then defined as:
\begin{equation}
\begin{aligned}
\frac{\partial f}{\partial x} (x) &:= \frac{1}{2} \left( \frac{\partial f}{\partial x_r} (x_r,x_i) - i \frac{\partial f}{\partial x_i} (x_r,x_i) \right), \\
\frac{\partial f}{\partial x^*} (x) &:= \frac{1}{2} \left( \frac{\partial f}{\partial x_r} (x_r,x_i) + i \frac{\partial f}{\partial x_i} (x_r,x_i) \right).
\end{aligned}
\label{eq:wirtinger_deriv}
\end{equation}
In practice, computing the derivative of $f$ with respect to $x$ (resp. $x^*$) can be done using usual differentiation by treating $x$ (resp. $x^*$) as a real variable with $x^*$ (resp. $x$) treated as a constant~\cite{Bouboulis2010,KreutzDelgado2005}:
\begin{align}
\label{eq:xconst}
\frac{\partial f}{\partial x} &= \frac{\partial f (x,x^*)}{\partial x}  \bigg\rvert_{x^* = \text{const.}} , \\
\frac{\partial f}{\partial x^*} &= \frac{\partial f (x,x^*)}{\partial x^*}  \bigg\rvert_{x = \text{const.}} .
\end{align}
Besides, if $f$ is real-valued, the following property is verified:
\begin{equation}
    \left ( \frac{\partial f}{\partial x}\right )^* = \frac{\partial f}{\partial x^*}.
    \label{eq:wirtinger_deriv_conjug}
\end{equation}
In a multivariate setting, the gradient of $f$ is then defined as:
\begin{equation}
    \nabla f = \left[ \frac{\partial f}{\partial x(1)}, \ldots ,\frac{\partial f}{\partial x(K)}  \right]^\mathsf H.
\end{equation}
When $f$ is additionally real-valued, the following property holds from~\eqref{eq:wirtinger_deriv} and~\eqref{eq:wirtinger_deriv_conjug}:
\begin{equation}
    \nabla_{\mathbb{R} } f := \left[ \frac{\partial f}{\partial x_r(1)}, \ldots ,\frac{\partial f}{\partial x_r(K)}  \right]^\mathsf H = 2 \mathfrak{R} ( \nabla f).
    \label{eq:grad_wirtinger_real}
\end{equation}

As an illustrative example, we derive the expression of the gradient in the Wirtinger Flow algorithm~\cite{WF}. The loss is:
\begin{equation}
    E (\mathbf x) = \frac{1}{2} \||\mathbf A \mathbf x|^2 - \mathbf r\|^2_2.
\end{equation}
Applying the chain rule yields:
\begin{equation}
    \nabla E (\mathbf{x}) = \big ( \nabla (|\mathbf A \mathbf x|^2 - \mathbf r)\big)^\mathsf H (|\mathbf A \mathbf x|^2 - \mathbf r) .
\end{equation}
Treating $\ve{x}^*$ as a constant like in~\eqref{eq:xconst}, the first term is given by:
\begin{align}
    \nabla (|\mathbf A \mathbf x|^2 - \mathbf r) &= \nabla (|\mathbf A \mathbf x|^2) \\
    &= \nabla ((\mathbf A \mathbf x)^* \odot (\mathbf A \mathbf x)) \\
    &= \text{diag}(\mathbf A \mathbf x)^* \nabla (\mathbf A \mathbf x) + \text{diag}(\mathbf A \mathbf x) \nabla \big ((\mathbf A \mathbf x)^*\big) \\
    &= \text{diag}(\mathbf A \mathbf x)^* \nabla (\mathbf A \mathbf x) + 0 \\
    &= \text{diag}(\mathbf A \mathbf x)^* \mathbf A.
\end{align}
We finally obtain:
\begin{align}
    \nabla E (\mathbf{x})  &= \mathbf A^\mathsf H \text{diag}(\mathbf A \mathbf x) (|\mathbf A \mathbf x|^2 - \mathbf r) \\
    &= \mathbf A^\mathsf H [(\mathbf A \mathbf x)\odot (|\mathbf A \mathbf x|^2 - \mathbf r)].
\end{align}

\section{Proximal operators}
\label{sec:apdx_prox}
\subsection{Definition}

The proximal operator of a convex function ${f : \mathbb R ^K \rightarrow \mathbb R \cup \{+\infty\}}$ is the operator mapping a vector $\mathbf y \in \mathbb R^K$ to the set of solutions of the following penalized optimization problem~\cite{combettes2011proximal}:
\begin{equation}
    \text{prox}_{\rho^{-1}f}(\mathbf y):=\underset{\mathbf x \in \mathbb R^K}{\text{argmin}}\quad f(\mathbf x)+\frac{\rho}{2}\|\mathbf x-\mathbf y\|_2^2.
    \label{eq:prox}
\end{equation}

\subsection{Proximal operator of usual Bregman divergences}

\begin{table}[t]
    \centering
    \caption{Proximal operators of some standard (convex) Bregman divergences. $\mathcal W$ is the Lambert W function (i.e., the inverse relation of $z \mapsto z e^z $) applied entry-wise.}
    \label{tab:prox}
    \normalsize
    \begin{tabular}{c|cc} \hline \hline
        Divergence & Expression & Proximal operator \\ \hline \hline
        Quadratic & $\frac{1}{2\rho} \| \cdot - \mathbf r \|^2_2$ & $\displaystyle \frac{\rho \mathbf y +  \mathbf r}{\rho  +1}$ \\
        KL right & $\rho^{-1}\mathcal D_{KL}(\mathbf r\,\bm|\,\cdot)$  & $\frac{1}{2\rho}(\mathbf y - 1 \pm \sqrt \Delta)$\\
        & &\small{with $\Delta := 4\rho\mathbf r+(1-\mathbf y)^2$}\\
        KL left & $\rho^{-1}\mathcal D_{KL}( \cdot\,\bm|\,\mathbf r)$ & $\rho^{-1} \mathcal W (\rho \mathbf r \odot e^{\rho \mathbf y})$\\
        IS left & $\rho^{-1}\mathcal D_{IS}(\cdot\,\bm|\,\mathbf r)$ &$\frac{1}{2\rho}(-\mathbf r^{-1}+\rho \mathbf y \pm \sqrt{\Delta'})$\\
        & &\small{with $\Delta' := 4\rho+(\mathbf r^{-1}-\rho\mathbf y)^2$}\\\hline \hline
    \end{tabular}
\end{table}

A closed-form expression of the proximal operator can be obtained for some of the usual Bregman divergences, such as the quadratic distance and the KL right and left divergences~\cite{combettes2011proximal, el2017proximity}. These are summarized in Table~\ref{tab:prox}.

To the best of our knowledge, the proximal operator of the IS divergence has not been derived in closed-form in the literature. Therefore, for the sake of completeness, we derive it hereafter.
Let us consider $\psi$ such that $\psi(z)=-\log z$. We consider the problem~\eqref{eq:prox} with $f(\mathbf x) = D_\psi(\mathbf x \,\bm |\,\mathbf r)$. Note that such a function is defined only for vectors with nonnegative entries. However, we can extend its definition domain to $\mathbb R^K$ by considering that $\mathcal D_\psi(\mathbf x \,\bm |\,\mathbf r) = + \infty$ if $\mathbf x \notin \mathbb R_+^K$~\cite{el2017proximity}.
We then search for $\mathbf x$ such that $\nabla Q(\mathbf x) = \bm 0$, where $Q(\mathbf x)=D_\psi(\mathbf x \,\bm |\,\mathbf r) + \frac \rho 2 \|\mathbf x - \mathbf y\|_2^2$.
We have:
\begin{align}
    \nabla Q(\mathbf x) &= \psi'(\mathbf x) - \psi'(\mathbf r) + \rho(\mathbf x - \mathbf y)
 \\ &= \mathbf r^{-1} - \mathbf x^{-1} + \rho(\mathbf x - \mathbf y).
\end{align}
Therefore,
\begin{align}
    \nabla Q(\mathbf x)=\bm 0 &\Longleftrightarrow \mathbf x \odot \mathbf r^{-1} - \bm 1 + \rho \mathbf x \odot (\mathbf x - \mathbf y) = \bm 0 \\ &\Longleftrightarrow \rho \mathbf x^2 + (\mathbf r^{-1}-\rho \mathbf y) \odot \mathbf x - \bm 1 = \bm 0.
\end{align}
Finally:
\begin{equation}
    \text{prox}_{\rho^{-1}\mathcal D_\psi( \cdot \,\bm |\,\mathbf r)}(\mathbf y) = \frac{1}{2\rho}(-\mathbf r^{-1}+\rho \mathbf y \pm \sqrt{\Delta'}),
\end{equation}
where $\Delta' := 4\rho+(\mathbf r^{-1}-\rho\mathbf y)^2$. Note that this operator might not be defined when one entry of $\mathbf{r}$ or $\mathbf{x}$ is null. To alleviate this problem in practice, we add a small value $\epsilon \ll 1$ to these vectors, as detailed in Appendix~\ref{apdx:regrad} for the gradient algorithm.

\subsection{Nonnegativity constraint in problem~\eqref{eq:ADMM_u_theta}}
\label{sec:apdx_prox_u}
Here we prove that the nonnegativity constraint on $\mathbf u$ in problem~\eqref{eq:ADMM_u_theta} can be ignored. Let us first rewrite this problem into scalar form, as this problem is separable entrywise:
\begin{equation}
\label{eq:amin_prox_right}
    \underset{u(k) \geq 0}{\text{argmin}}\, d_\psi (r(k) \, \bm | \, u(k)) + \frac \rho 2 \|| h(k)|- u(k)\|^2.
\end{equation}
We will remove the index $k$ in what follows for clarity. We aim to prove that:
\begin{equation}
\label{eq:scal_ineq}
    \mbox{If } u<0 \mbox{,\quad}
    d_\psi(r\,\bm|\,0) + \frac \rho 2 |h|^2\,\leq\, d_\psi(r\,\bm|\,u) + \frac \rho 2 ||h|-u|^2,
\end{equation}
If this inequality holds, then the minimizer of the function defined in~\eqref{eq:amin_prox_right} necessarily belongs to $\mathbb{R}_+$. Consequently, the nonnegativity constraint can be dismissed. Equation~\eqref{eq:scal_ineq} rewrites:
\begin{equation}
    \psi(r) - \psi(0) - \psi'(0)r + \frac \rho 2 |h|^2 \leq \psi(r) - \psi(u)  - \psi'(u)(r-u) + \frac \rho 2 ||h|-u|^2,
\end{equation}
which is equivalent to:
\begin{equation}
    \psi(0) - \psi(u) + r \psi'(0) - \psi'(u)(0-u)  - r \psi'(u) + \frac \rho 2 [-2u|h|+u^2]\geq 0,
\end{equation}
which finally rewrites:
\begin{equation}
     \underbrace{d_\psi(0\,\bm|\,u)}_{\text{term }1}+ \underbrace{r(\psi'(0) - \psi'(u))}_{\text{term }2} + \underbrace{\frac \rho 2 [-2u|h|+u^2]}_{\text{term }3}\geq 0.
\end{equation}
The latter inequality holds for the following reasons:
\begin{itemize}
    \item Term $1$ is nonnegative by nonnegativity of Bregman divergences.
    \item Term $2$ is nonnegative by convexity of $\psi$ and nonnegativity of $r$:
    $\psi$ is convex, therefore $\psi'$ is monotonically non-decreasing. As
    $u<0$, $\psi'(u) \leq \psi'(0)$ and $r(\psi'(0) - \psi'(u))\geq 0$.
    \item Term $3$ is nonnegative because $u$ is negative.
\end{itemize}
Therefore,~\eqref{eq:scal_ineq} holds, which demonstrates that the nonnegativity constraint in~\eqref{eq:ADMM_u_theta} can be dismissed. Finally, using a similar proof, we can show that the same holds for the ``left" PR problem.

\section{Algorithms derivations for real-valued signals}
\label{sec:apdx_real}

We here discuss the adaptation of our proposed gradient and ADMM algorithms to the specific case when the input signal is real-valued $\mathbf{x} \in \mathbb{R}^L$.

In this setting, the gradient algorithm can be easily deduced from its complex-valued counterpart. Indeed, since $\mathbf{x}$ is real-valued, the gradient of $J$ simply reduces to $\nabla_{\mathbb{R}} J( \mathbf x)$, as defined in Appendix~\ref{sec:apdx_wirtinger}. According to the property~\eqref{eq:grad_wirtinger_real}, this gradient is given by:
\begin{equation}
    \nabla_{\mathbb{R}} J( \mathbf x) = 2 \mathfrak{R} (  \nabla J( \mathbf x) ).
    \label{eq:nabla_J}
\end{equation}
where $ \nabla J( \mathbf x)$ is computed using the Wirtinger derivatives. Consequently, the gradient update rule is similar to the complex-valued case, up to a constant factor of $2$ and with the difference that we only need to retain the real part after applying $\mathbf{A}^\mathsf{H}$ (in practice, the inverse STFT).

Regarding the ADMM algorithm, we need to address the following sub-problem, in lieu of~\eqref{eq:admm_prob_x_complex}:
\begin{equation}
    \underset{\mathbf x \in \mathbb{R}^L }{\text{min}} \, || (\mathbf A\mathbf x)^d - \mathbf{b} ||_2^2.
    \label{eq:prob_admm_x}
\end{equation}
where we note $\mathbf{b} =  \mathbf u_{t+1} \odot e^{i \bm \theta_{t+1}} - \frac{\bm \lambda_t}{\rho}$. Since we only use ADMM algorithms with $d=1$ in our experiments, we focus hereafter on this setting. By using again~\eqref{eq:grad_wirtinger_real}, we compute the gradient of the loss in~\eqref{eq:prob_admm_x} and set it at $0$:
\begin{equation}
     2 \mathfrak{R} ( \mathbf{A}^{\mathsf H} \mathbf A \mathbf x -  \mathbf{A}^{\mathsf H} \mathbf b )   = 0.
\end{equation}
This yields the following solution:
\begin{equation}
     \mathbf x = (\mathfrak{R} (\mathbf{A}^{\mathsf H} \mathbf A))^{-1} \mathfrak{R}( \mathbf{A}^{\mathsf H} \mathbf b).
     \label{eq:up_x_admm_R_alt}
\end{equation}
When using the STFT with a self-dual window we have  $\mathbf{A}^{\mathsf H}\mathbf{A} = \ve{I}_L$ and the update becomes
\begin{equation}
     \mathbf x = \mathfrak{R} ( \mathbf{A}^{\mathsf H} \mathbf b ).
\end{equation}
It is the same update as in the complex-valued case~\eqref{eq:admm_x} up to retaining the real part after applying the inverse STFT $\ve{A}^\mathsf H$.

\section{Regularized gradient expression}
\label{apdx:regrad}
For some Bregman divergences and/or exponents $d$, the gradient of the loss functions~\eqref{eq:bregpr_right} and~\eqref{eq:bregpr_left} is not defined when one or more coefficients of $\mathbf A\mathbf x$ are zero-valued, which leads to division by zero and other potential numerical or conceptual issues. This is the case, for instance, when $d\le 1$, when computing $|\mathbf A \mathbf x|^{d-2}$ with $d<2$, or when computing $\psi'(|\mathbf A \mathbf x|^d)$ for a beta-divergence such that $\beta\leq1$. Therefore, we propose a rigorous treatment of this issue by considering regularized losses. More specifically, we consider the following alternative loss for the PR right problem (a similar technique is used for treating its left counterpart):
\begin{equation}
  J_\varepsilon(\mathbf x) := \mathcal D_\psi \left ( (\mathbf r^{\frac 2 d}+\varepsilon)^{\frac d 2}\, \bm | \, (|\mathbf A \mathbf x|^2+\varepsilon)^{\frac d 2}\right ),
\end{equation}
with $\varepsilon \ll 1$, such that $J_\varepsilon$ is now always defined and differentiable at $0$. This yields the corresponding regularized gradient expression:
\begin{equation}
  \nabla J_\varepsilon (\mathbf x)=\frac d 2 \mathbf A^\mathsf H\left [ (|\mathbf A \mathbf x|^2+\varepsilon)^{\frac d 2 -1} \odot \mathbf A \mathbf x \odot \mathbf{g}_{\psi, \varepsilon} \right],
\end{equation}
with
\begin{equation}
  \mathbf{g}_{\psi, \varepsilon} =
        \psi''((|\mathbf A \mathbf x|^2+\varepsilon)^{\frac d 2})\odot((|\mathbf A \mathbf x|^2+\varepsilon)^{\frac d 2} -(\mathbf r^{\frac 2 d}+\varepsilon)^{\frac d 2}).
\end{equation}
For the PR left problem, a similar expression is obtained:
\begin{equation}
  \mathbf{g}_{\psi, \varepsilon} =
        \psi'((|\mathbf A \mathbf x|^2+\varepsilon)^{\frac d 2}) - \psi'((\mathbf r^{\frac 2 d}+\varepsilon)^{\frac d 2}).
\end{equation}
We used this variant in our experiments, and implemented it in practice with $\varepsilon = 10^{-8}$.

\section*{Acknowledgments}

The authors would like to thank the guest editors and the anonymous reviewers for their thorough reading and valuable comments, which helped significantly improving this paper.

\bibliographystyle{IEEEtran}
\bibliography{biblio}

%








\end{document}